\documentclass[preprint,3p,times,twocolumn]{elsarticle}
\pdfoutput=1

\usepackage{hhline}
\usepackage{float}
\usepackage{amsmath}
\usepackage{amsthm}
\usepackage{amsfonts}
\usepackage{amssymb}
\usepackage{listings}
\usepackage{caption}
\usepackage{subcaption}
\usepackage{tabularx}
\usepackage[usenames,dvipsnames,svgnames,table]{xcolor}
\usepackage{tikz}
\usetikzlibrary{shapes,shapes.geometric,arrows,fit,calc,positioning,automata,}

\usepackage{paralist}
\usepackage{enumitem}
\usepackage{framed}
\usepackage[capitalize,nameinlink]{cleveref}

\usepackage[Algorithm]{algorithm}
\usepackage[noend]{algpseudocode} % http://ctan.org/pkg/algorithmicx
\algrenewcommand\alglinenumber[1]{\tiny #1:}

\usepackage{forest}
\usepackage{smartdiagram}
\usepackage{multirow}
\usesmartdiagramlibrary{additions} % required in the preamble
\usetikzlibrary{arrows} % required in the preamble

\tikzset{
	vertex style/.style={
		draw=#1,
		thick,
		fill=#1!50,
		text=black,
		ellipse,
		minimum width=0.2cm,
		minimum height=0.2cm,
		font=\tiny\sffamily,
		inner sep=2pt,
	},
	vertexr style/.style={
		draw=#1,
		thick,
		fill=#1!50,
		text=black,
		rectangle,
		minimum width=0.2cm,
		minimum height=0.2cm,
		font=\tiny\sffamily,
		inner sep=2pt,
	},
	text style/.style={
		sloped, % the text will be parallel to the connection 
		text=black,
		font=\tiny\sffamily,
		above
	},
	textb style/.style={
		sloped, % the text will be parallel to the connection 
		text=black,
		font=\tiny\sffamily,
		below
	}
}

\algrenewcommand\algorithmicindent{0.05cm}%

\lstdefinestyle{tiny}{
	basicstyle=\ttfamily\tiny, 
	numbers=none
}

\lstdefinestyle{script}{
	basicstyle=\ttfamily\scriptsize, 
	numbers=none
}

\lstdefinelanguage{sparql}{
	morecomment=[l][\color{teal}]{\#},
	morestring=[b][\color{blue}]\",
	morekeywords={SELECT,CONSTRUCT,DESCRIBE,ASK,WHERE,FROM,NAMED,PREFIX,BASE,OPTIONAL,FILTER,GRAPH,LIMIT,OFFSET,SERVICE,UNION,EXISTS,NOT,BINDINGS,MINUS,a,as,GROUP,BY,SUM,AVG,VALUES},
	sensitive=false
}

\lstdefinelanguage{ql}{
	morecomment=[l][\color{teal}]{\#},
	morestring=[b][\color{blue}]\",
	morekeywords={ROLLUP,DRILLDOWN,DICE,SLICE,OR,AND,NOT},
	sensitive=true
}

\lstdefinestyle{sparql}{
	basicstyle=\ttfamily\scriptsize, 
	language=sparql,
	columns=fullflexible, 
	numberstyle= \tiny, %\scriptsize,
	numbers=left,
	frame=lines,
	tabsize=2}

\lstdefinestyle{qlquery}{
	basicstyle=\ttfamily\scriptsize,
	language=ql,
	numbers=none,
	frame=none,
	tabsize=2}

\theoremstyle{plain}
\newtheorem{definition}{Definition}[section]
\newtheorem{property}{Property}[section]
\theoremstyle{definition}
\newtheorem{example}{Example}[section]
\theoremstyle{remark}
\newtheorem{remark}{Remark}

\newenvironment{querynonum}{\sffamily \scriptsize \framed 
	\noindent\ignorespaces} {\ignorespacesafterend\endframed}

\def\QL{{{CQL}}} 
\newcommand {\te}[1]{{\small  \textsf{#1}}}
\newcommand {\ttt}[1]{{\small  \texttt{#1}}}
	
\newcolumntype{L}[1]{>{\raggedright\let\newline\\\arraybackslash\hspace{0pt}}m{#1}}
\newcolumntype{C}[1]{>{\centering\let\newline\\\arraybackslash\hspace{0pt}}m{#1}}
\newcolumntype{R}[1]{>{\raggedleft\let\newline\\\arraybackslash\hspace{0pt}}m{#1}}
\newcolumntype{Y}{>{\centering\arraybackslash}X}

\begin{document}

\begin{frontmatter}

\title{Efficient Analytical Queries on Semantic Web Data Cubes}

\author[um]{Lorena Etcheverry}
\cortext[mycorrespondingauthor]{Corresponding author}
\ead{lorenae@fing.edu.uy}

\author[iba]{Alejandro A. Vaisman}
\address[um]{Instituto de Computaci\'{o}n, Facultad de Ingenier\'{i}a, UdelaR, Ave Julio Herrera y Reissig 565, Montevideo, Uruguay}
\address[iba]{Instituto Tecnol\'{o}gico de Buenos Aires, 25 de Mayo 457, Buenos Aires, Argentina}

\begin{abstract}
  
The amount of multidimensional data published on the semantic web (SW) is constantly increasing,  due to initiatives such as Open Data and Open Government Data, among other ones. Models, languages, and  tools,  that  allow obtaining valuable information efficiently, are thus required. Multidimensional data are  typically represented as data cubes, and exploited using Online Analytical Processing (OLAP) techniques. The  RDF Data Cube Vocabulary, also 
denoted QB, is the current W3C standard to represent statistical data on the SW.
Given that QB does not include key features needed for OLAP analysis, 
in previous work we have proposed an extension, denoted QB4OLAP, to overcome this problem without the need of modifying already published data. 

Once data cubes are appropriately represented on the SW, we need mechanisms to analyze them. However, in the current state-of-the-art, writing efficient analytical queries over SW data cubes demands a deep knowledge of standards like  RDF and SPARQL. These  skills are  unlikely to be found in typical analytical users. Further, OLAP languages like MDX are far from being easily understood by the final user. The lack of friendly tools to exploit multidimensional data on the SW is a barrier that needs to be broken to promote the publication of such data. This is the problem we address in this paper. Our approach is based on allowing analytical users to write queries using what they know best: OLAP operations over data cubes, without dealing with SW technicalities. For this, we devised CQL (standing for Cube Query Language), a simple, high-level query language that operates  over data cubes.  Taking advantage of structural metadata provided by QB4OLAP, we  translate CQL queries into SPARQL ones. Then, we propose query improvement strategies to produce efficient SPARQL queries, adapting general-purpose SPARQL query optimization techniques. We evaluate our implementation using the Star-Schema benchmark, showing that our proposal outperforms others. The \textit{QB4OLAP toolkit}, a web application that allows exploring and  querying (using CQL) SW data cubes,  completes our contributions.  
\end{abstract}


\end{frontmatter}

%\linenumbers

%\input{intro}
%\input{examples}
%\input{qb4olap2}
%\input{querying}
%\input{implementation}
%\input{evaluation}
%\input{related}
%\input{conclusions}
%
%\appendix
%\input{appendix}
%\input{ack}

\section{Introduction}
\label{sec:intro}  

Data Warehouses (DW) integrate  multiple data sources  for 
analysis and decision support, representing data according to the 
Multidimensional (MD) Model.
This model organizes data in MD data cubes, where hierarchical {\em dimensions} represent the perspectives that characterize  {\em facts}. The latter are usually associated with quantitative data, also known as \emph{measures}. 
Data cube measures can be aggregated, disaggregated, and filtered  
using dimensions, and this process is called Online Analytical Processing 
(OLAP). 

DW and OLAP  had been typically used as techniques for data analysis 
\textit{within} organizations, based on high quality internal data, and mostly 
using   commercial tools with proprietary formats.  
However, initiatives such as Open Data\footnote{\url{http://okfn.org/opendata/}} 
and Open Government Data\footnote{\url{http://opengovdata.org/}} 
are encouraging organizations to publish and share MD data on the web. 
In addition, the \textit{Linked Data} (LD) paradigm promotes  a set of best practices for publishing and 
interlinking structured data on the web, using  standards, like RDF\footnote{\url{https://w3.org/RDF/}}, 
and SPARQL.\footnote{\url{http://w3.org/TR/sparql11-query/}} 
At the time of writing this paper, the amount of open data available 
as LD is approximately 90 billion triples in over 3,300 data sets, most 
of them freely accessible via SPARQL query 
endpoints.\footnote{\url{http://stats.lod2.eu/}}
However, LD recommendations focus on the representation  of
relational data, but they are insufficient to represent other data models, in 
particular MD data.

In this new context, the Business Intelligence (BI) community faces several challenges. 
First, there is a need for instruments to represent MD data and metadata (e.g., dimensional structure, which  is essential to adequately interpret and reuse data) 
using Semantic Web (SW) standards. 
Second, it is necessary to provide mechanisms to analyze SW data \textit{\'{a} la} OLAP.  
Regarding the first challenge, the \textit{RDF Data Cube Vocabulary}\cite{Cyganiak2014} 
(QB) is the current W3C standard to represent statistical data following LD principles.
There is  already a considerable number of data sets published using QB. 
However, this vocabulary does not include key features needed for OLAP analysis,
 like dimensional hierarchies and aggregate 
functions. To address this problem, in previous work, we have proposed a new vocabulary called QB4OLAP~\cite{Etcheverry2012a,Etcheverry2012b}, 
which extends QB in order to overcome these limitations.
 QB4OLAP also allows reusing data already published in QB, 
just by adding the needed MD schema semantics, and the corresponding data  
instances.    

 The work we present in this paper is aimed at tackling the second challenge above.
 To this end, we propose a high-level query language for OLAP, denoted \QL, where
 the \textit{main data type is the data cube}. Our approach is based on a
 clear separation between the conceptual and the logical levels, a feature 
 that is not common in traditional OLAP systems,where popular OLAP query and 
 analysis languages, such as 
 MDX,\footnote{\url{http://microsoft.com/msj/0899/mdx/mdx.aspx}} 
 operate  at  the  logical  level  and  require,  in  order  to  be  able  to write    
 queries, the user's deep understanding of how data are actually stored~\cite{CCG+13}.  
 To achieve this separation, we start defining a data model for MD data 
 cubes, and an algebra (which is a subset of the 
 so-called Cube Algebra proposed in~\cite{CCG+13}), 
 composed of  a collection of 
 operators, with a clearly defined  semantics. 
 This  algebra will be the basis of our high-level OLAP query language, denoted 
 \textit{\QL} (standing for \textit{Cube Query Language}), and is composed of a 
 collection of operations that manipulate a data cube, which is the only kind 
 of object that the user will be aware of. The user will thus write her  
 queries at the conceptual level  using \QL, and we provide mechanisms to 
 translate these queries into  
 SPARQL ones, over the QB4OLAP-based RDF representation (at the logical level). 
 The main advantage of this approach is that it allows users to perform OLAP queries   \textit{directly} over QB4OLAP cubes on the SW, without dealing with RDF or SPARQL technicalities. Note that, in general, OLAP users know how to manipulate a data cube through the
typical roll-up, drill-down, and slice-dice operations, but it is unlikely 
that they would be  familiar with SPARQL or the SW.   
Also, SPARQL optimization tips and best practices could  be incorporated into 
the~\QL~to SPARQL translation process, to produce efficient  queries, not an easy task for an average user. On the other hand, SW users know SPARQL and RDF very well, but the cube metaphor may help them to perform analytical queries easier and more intuitively than operating directly over the RDF representation.

More concretely, as our \textit{first contribution}, we present  a data 
model for OLAP and propose an algebra  and a high-level query language 
based on it, namely~\QL, where  the main 
data type is the data cube.  The semantics of the algebra operators is clearly 
defined using the notion of a \textit{lattice of cuboids}, which is 
used for query processing and rewriting. 
  
The core of this paper is about automatically producing 
an efficient SPARQL implementation of \QL~queries over QB4OLAP data cubes. 
Thus, as our \textit{second} and
\textit{main contribution} we: (1) 
 Present a high-level heuristic query simplification strategy for \QL; (2)
 Propose  algorithms to automatically translate \QL~queries into equivalent  
SPARQL ones over QB4OLAP data cubes; (3) 
 Propose a heuristic-based strategy to improve the performance of the    
 SPARQL queries produced in (2); (4) Introduce a benchmark, based on TPC-H 
  and the Star-Schema benchmark, to evaluate the performance of the   SPARQL 
  queries; we show that  our improvement procedure substantially  
  speeds-up    the query evaluation process, and outperforms other 
  proposals; (5) Present the \textit{QB4OLAP toolkit}, a web application that 
allows exploring and querying QB4OLAP cubes.
   
The remainder of this paper is organized as follows. 
Section \ref{sec:examples} presents the 
running example we will use in this work.
Section \ref{sec:qb4olap} briefly sketches the QB4OLAP vocabulary.
Section \ref{sec:queries} presents our approach to querying QB4OLAP data cubes. 
In Section ~\ref{sec:impl} we concisely present our implementation, while Section 
~\ref{sec:eval} reports our experimental results. 
Section~\ref{sec:related} discusses related work. We conclude in 
Section~\ref{sec:conclusion}. 

\begin{remark}
Our  proposal for querying QB4OLAP data cubes has been previously briefly sketched  
in~\cite{Etcheverry2016}, while in this paper we develop those ideas in-depth, and provide a detailed experimental study, not included in previous work.
\end{remark}

\section{Running example}
\label{sec:examples}

Throughout this paper we use an example based on statistical data about
asylum  applications to the European Union, provided by  
Eurostat.\footnote{\url{http://ec.europa.eu/eurostat/web/products-datasets/-/migr_asyappctzm}}  
This data set contains information about the number of asylum applicants per 
month, age, sex, citizenship, application type, and country that receives the application.  It is published in the Eurostat LD dataspace,\footnote{\url{http://eurostat.linked-statistics.org/}}  using the QB vocabulary. 
QB data sets are composed of a set of \textit{observations} representing 
data instances according to a \textit{data structure definition}, which 
describes the schema of the data cube. 
We  enriched the original data set in order to enhance the analysis possibilities. Making use of the features of QB4OLAP, we were able to reuse the published observations, so we only created new dimensions, and represented them using QB4OLAP structural metadata.
%, as we discuss later. 

\begin{figure}[!t]
\centering
\includegraphics[width=\linewidth]{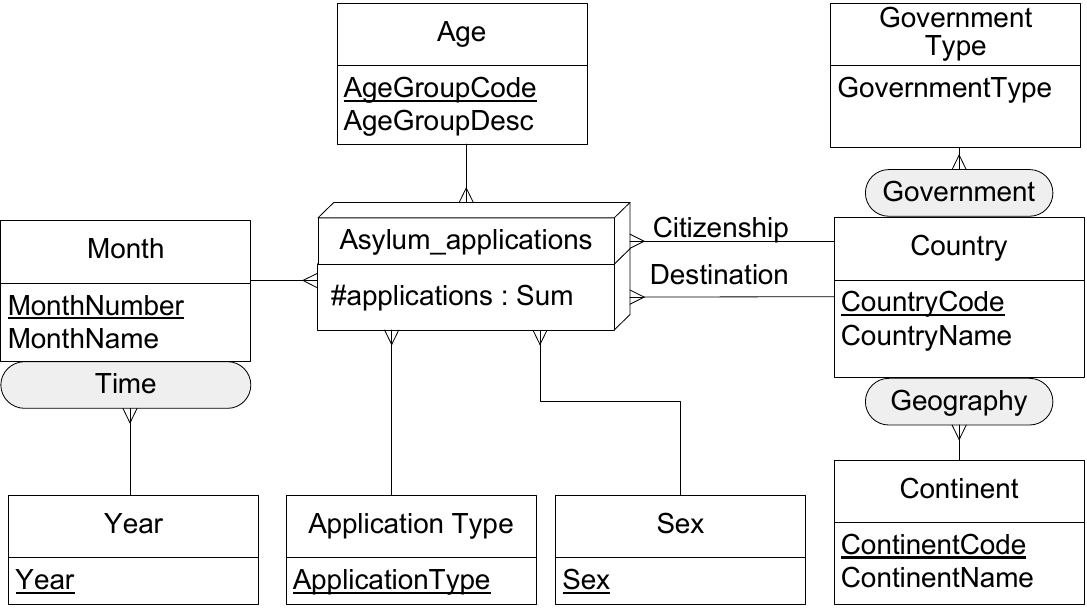}
\caption {Conceptual schema of the asylum applications cube}
\label{fig.concEuro}
\end{figure}

Figure~\ref{fig.concEuro} shows the resulting conceptual 
schema of the  data cube, using the MultiDim notation \cite{VZ14}. 
The \te{asylum\_applications} fact contains  a measure
(\te{\#applications}) that represents the number of applications. 
This measure can be analyzed according to six analysis dimensions:  
\te{sex}  of the applicant, \te{age} which organizes applicants according to their 
age group, \te{time} which represents the time of the application and consists of two 
levels (\te{month} and \te{year}), \te{application\_type} that tells if the 
applicant is a first-time applicant 
or a returning one, and a geographical dimension that organizes countries into continents (\te{Geography} hierarchy) or according to its government type (\te{Government} hierarchy). 
This geographical dimension participates in the cube with two different roles: the \te{citizenship} of the asylum applicant, 
and the \te{destination} country of the application. 
To create these hierarchies, we  enriched the existent data set with 
DBpedia\footnote{\url{http://dbpedia.org}} data,  retrieving, for each country, its government type,
and the continent it belongs to. 

As an example, Table \ref{fig.cubeinstance} shows some observations in tabular format . The first row 
lists the dimensions in the cube, and the second row lists the dimension level that corresponds to the observations.

\begin{table}[h]
\caption{Tabular representation of sample observations in the asylum applications datacube.}
\tiny
\begin{tabularx}{\linewidth}{@{\extracolsep{\fill}} |Y|C{.6cm}|C{.8cm}|C{.8cm}|C{.8cm}|C{.8cm}|C{.8cm}|}
	\hline
	\textbf{Sex} & \textbf{Age} & \textbf{Time} &\textbf{Application type} & 
	\textbf{Citizenship} & \textbf{Destination} & \textbf{Measures} \\ \hline
	\textit{Sex} & \textit{Age} & \textit{Month} & \textit{Application type} & 
	\textit{Country} & \textit{Country} &\textit{ \#applications} \\ \hline
	F & 18 to 34 & 201409, September 2014 & new applicant & SY, Syria & DE, Germany & 425 \\ \hline
	M & 18 to 34 & 201409, September 2014 & new applicant & SY, Syria & DE, Germany & 1680 \\ \hline
	M & 18 to 34 & 201409, September 2014 & new applicant & SY, Syria & FR, France & 95 \\ \hline
\end{tabularx}
\label{fig.cubeinstance}
\end{table}

Over the new cube, depicted in Figure~\ref{fig.concEuro}, we can pose queries like ``\textit{Total asylum applications per year}'',  or 
``\textit{Total asylum applications per year submitted by Asian citizens to France or United Kingdom, where this number is higher than 5,000}'', which we discuss later in this paper.

\section{The QB4OLAP vocabulary}
\label{sec:qb4olap}

In QB, the schema of a data set is specified by means of the \textit{data structure definition} (DSD), an instance of the class  \ttt{qb:DataStructureDefinition}. 
This specification is formed by \textit{components}, which represent \textit{dimensions}, \textit{measures}, and \textit{attributes}.  
\textit{Observations} (in OLAP terminology, \textit{fact instances}) represent points in a  MD data space indexed by \textit{dimensions}. 
These points are modelled using instances of the class  \ttt{qb:Observation}, and are organized in \textit{data sets}, defined as instances of the class 
\ttt{qb:DataSet}, where each data set is associated with a DSD that describes the structure of a cube. 
Finally, each observation is linked to a member in each dimension of the corresponding DSD via instances of the class \ttt{qb:DimensionProperty}; analogously, each observation is  associated with measure values via instances of the class \ttt{qb:MeasureProperty}.

The QB4OLAP\footnote{\url{http://purl.org/qb4olap/cubes}} vocabulary extends QB to allow  representing the most common features of the MD model. In this way,  we can represent a dimension schema as composed of hierarchies of aggregation levels. 
We can also represent the allowed aggregate functions, rollup relationships (i.e., 
the parent-child relationships between dimension level members), and the descriptive attributes of dimension levels. 
QB4OLAP allows   operating over observations  already published using QB, without the need  of rewriting them. 
This is relevant since  in a typical MD model, observations are the 
largest part of the data, while dimensions are usually orders of magnitude 
smaller. In this section we sketch the key aspects of the vocabulary, and refer the reader to~\cite{etcheverry2015,Vaisman2015} 
for details and a thorough comparison between QB and QB4OLAP. 
 
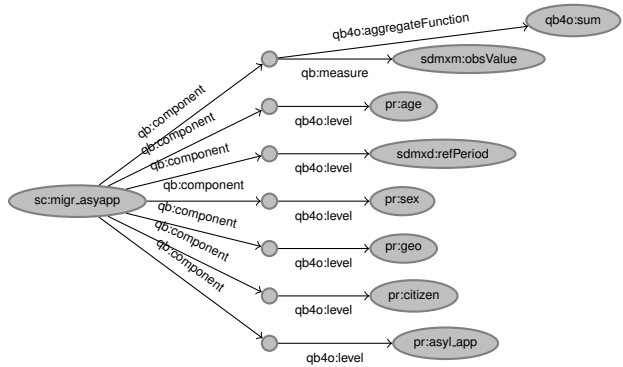
\begin{figure}[!t]
	
\begin{tikzpicture}
\node[vertex style=Gray] (cube) {sc:migr\_asyapp};
\node[vertex style=Gray, right=1.5 cm of cube] (b4) {}
edge [<-,black] node[text style]{qb:component} (cube);
\node[vertex style=Gray, above =.4cm of b4] (b3) {}
edge [<-,black] node[text style]{qb:component} (cube);
\node[vertex style=Gray, above =.4cm of b3] (b2) {}
edge [<-,black] node[text style]{qb:component} (cube);
	\node[vertex style=Gray, above =.4cm of b2] (b1) {}
	edge [<-,black] node[text style]{qb:component} (cube);

\node[vertex style=Gray, below =.4cm of b4] (b5) {}
edge [<-,black] node[text style]{qb:component} (cube);
\node[vertex style=Gray, below =.4cm of b5] (b6) {}
edge [<-,black] node[text style]{qb:component} (cube);
\node[vertex style=Gray, below =.4cm of b6] (b7) {}
edge [<-,black] node[text style]{qb:component} (cube);

\node[vertex style=Gray, right=1.5 cm of b1] (ms) {sdmxm:obsValue}
edge [<-,black] node[textb style]{qb:measure} (b1);
\node[vertex style=Gray, above right = .3cm of ms]  (af) {qb4o:sum}
edge [<-,black] node[text style]{qb4o:aggregateFunction} (b1);
\node[vertex style=Gray, right=1.2 cm of b2] (lv1) {pr:age}
edge [<-,black] node[textb style]{qb4o:level} (b2);
\node[vertex style=Gray, right=1.2 cm of b3] (lv2) {sdmxd:refPeriod}
edge [<-,black] node[textb style]{qb4o:level} (b3);
\node[vertex style=Gray, right=1.2 cm of b4] (lv3) {pr:sex}
edge [<-,black] node[textb style]{qb4o:level} (b4);
\node[vertex style=Gray, right=1.2 cm of b5] (lv4) {pr:geo}
edge [<-,black] node[textb style]{qb4o:level} (b5);
\node[vertex style=Gray, right=1.2 cm of b6] (lv5) {pr:citizen}
edge [<-,black] node[textb style]{qb4o:level} (b6);

\node[vertex style=Gray, right=1.5 cm of b7] (lv6) {pr:asyl\_app}
edge [<-,black] node[textb style]{qb4o:level} (b7);
\end{tikzpicture}
\caption{QB4OLAP representation of Asylum applications data cube schema.}
\label{fig:cubeSchema}
\end{figure}

In QB4OLAP, facts represent relationships 
\textit{between dimension levels}, and  observations (fact instances)  map 
\textit{level members} to measure values. 
Thus, QB4OLAP represents the structure of a data set in terms of 
\textit{dimension levels} and measures, instead of \textit{dimensions} and measures (which 
is the case of QB),  allowing us to specify data cubes at different granularity levels 
in  the cube dimensions.  
Accordingly, the schema of a cube in QB4OLAP is defined, like in QB, via a DSD, but in terms of dimension levels, introducing the
class \ttt{qb4o:LevelProperty}  to represent them.
QB4OLAP also introduces the class \ttt{qb4o:AggregateFunction} to represent the \textit{aggregate functions} that should be applied to summarize
measure values. The property  \ttt{qb4o:aggregateFunction}  associates  measures with  aggregate functions in the DSD.   
Figure \ref{fig:cubeSchema} shows an excerpt of the representation of the Asylum applications data cube schema using the QB4OLAP vocabulary. In the figure, empty circles represent blank nodes. The node labeled \ttt{sc:migr\_asyapp} represents the DSD of the cube.

Dimension hierarchies and levels are first-class citizens 
in a MD model for OLAP. Therefore, 
QB4OLAP focuses on their  representation, and several classes and properties are 
introduced for that. 
To represent \textit{dimension level attributes}, QB4OLAP provides  the class   \ttt{qb4o:LevelAttribute}, linked  to \ttt{qb4o:LevelProperty}  via the \ttt{qb4o:hasAttribute} 
property. 
The class \ttt{qb4o:Hierarchy} represents \textit{dimension hierarchies}, and the relationship between dimensions and hierarchies is represented via the property 
\ttt{qb4o:hasHierarchy} and its inverse \ttt{qb4o:inDimension}.
To support the fact that a level may belong to different hierarchies, and  each level may have a different set of parent levels, the concept of \ttt{qb4o:HierarchyStep} is introduced.
This represents the reification of the parent-child relationship between two levels. Hierarchy steps are implemented as blank nodes, and each hierarchy step is linked to its component 
levels using the \ttt{qb4o:childLevel} and \ttt{qb4o:parentLevel} properties, respectively. It is also 
associated with the hierarchy it belongs to, through  the property  \ttt{qb4o:inHierarchy}. 
Also, the property  \ttt{qb4o:pcCardinality} represents the cardinality of the relationships between level members in this step. 

In earlier versions of QB4OLAP, the \textit{rollup relationships} (in what follows, RUPs) between levels were represented, at the instance level, 
using the property \texttt{skos:broader}. Although this solution is enough for most kinds of MD hierarchies, it does not suffice to represent, 
at the instance level, dimensions with more than one RUP relationships (or functions) between the same pair of levels, usually denoted as  \textit{parallel dependent hierarchies} \cite{VZ14}. 
As an example, consider a geographical dimension with two levels: \textsf{Employee} and \textsf{City}. These levels participate in two hierarchies: one that represents 
the city where the employee lives (say, \textsf{LivesIn}), and another that represents the city where the employee works (\textsf{WorksIn}). 
It is easy to see that an Employee may live and work in different cities; in order to represent this at the instance level, we need to define two different RDF properties,
one for each RUP. Therefore, in QB4OLAP version 1.3 we introduced a mechanism to associate each hierarchy step with a user-defined property that implements the RUP at the instance level. 
These properties are instances of the class \ttt{qb4o:RollupProperty}, and are linked to each hierarchy step via the property \ttt{qb4o:rollup}. 

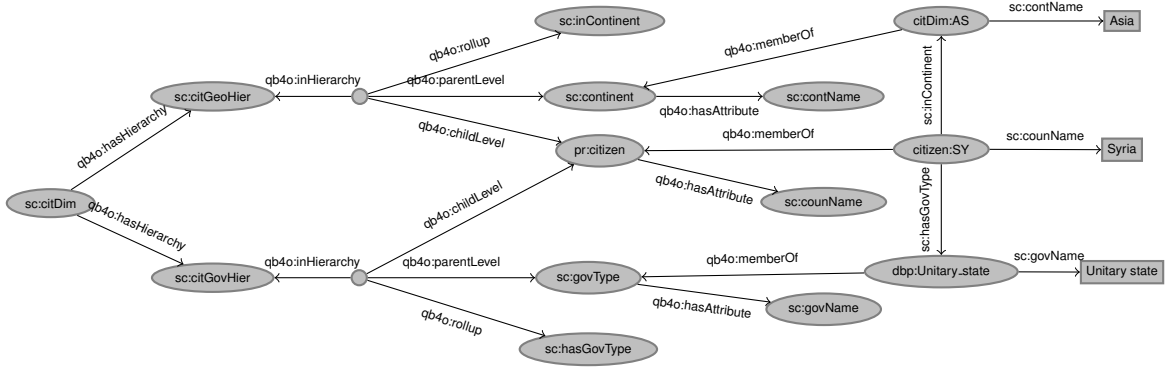
\begin{figure*}[!t]
\centering

\begin{tikzpicture}
\node[vertex style=Gray] (dim) {sc:citDim};
\node[vertex style=Gray, above right=1.6cm of dim] (h1) {sc:citGeoHier}
edge [<-,black] node[text style]{qb4o:hasHierarchy} (dim);
\node[vertex style=Gray, below =2cm of h1] (h2) {sc:citGovHier}
edge [<-,black] node[text style]{qb4o:hasHierarchy} (dim);

\node[vertex style=Gray, right=1 cm of h1] (s1) {}
edge [->,black] node[text style]{qb4o:inHierarchy} (h1);

\node[vertex style=Gray, right=2.3cm of s1] (lv2) {sc:continent}
edge [<-,black] node[text style]{qb4o:parentLevel} (s1);

\node[vertex style=Gray, right =1.4cm of lv2] (at2) {sc:contName}
edge [<-,black] node[textb style]{qb4o:hasAttribute} (lv2);

\node[vertex style=Gray, above of= lv2] (ru) {sc:inContinent}
edge [<-,black] node[text style]{qb4o:rollup} (s1);
\node[vertex style=Gray, right=3 cm of ru] (lv2m) {citDim:AS}
edge [->,black] node[text style]{qb4o:memberOf} (lv2);
\node[vertexr style=Gray, right=1.5 cm of lv2m] (av2) {Asia}
edge [<-,black] node[text style]{sc:contName} (lv2m);

\node[vertex style=Gray, below =.3cm of lv2] (lv1) {pr:citizen}
edge [<-,black] node[textb style]{qb4o:childLevel} (s1);

\node[vertex style=Gray, below=1cm of at2] (at1) {sc:counName}
edge [<-,black] node[textb style]{qb4o:hasAttribute} (lv1);

\node[vertex style=Gray, below=1.3cm of lv2m] (lv1m) {citizen:SY}
edge [->,black] node[text style]{qb4o:memberOf} (lv1)
edge [->,black] node[text style]{sc:inContinent} (lv2m);

\node[vertexr style=Gray, below=1.4cm of av2] (av1) {Syria}
edge [<-,black] node[text style]{sc:counName} (lv1m);

\node[vertex style=Gray, right=1 cm of h2] (s2) {}
edge [->,black] node[text style]{qb4o:inHierarchy} (h2)
edge [->,black] node[text style]{qb4o:childLevel} (lv1);

\node[vertex style=Gray, right =2.2cm of s2] (lv4) {sc:govType}
edge [<-,black] node[text style]{qb4o:parentLevel} (s2);
\node[vertex style=Gray, below=1cm of at1] (at3) {sc:govName}
edge [<-,black] node[textb style]{qb4o:hasAttribute} (lv4);
\node[vertex style=Gray, below =0.5cm of lv4] (ru2) {sc:hasGovType}
edge [<-,black] node[textb style]{qb4o:rollup} (s2);

\node[vertex style=Gray, below=1.2cm of lv1m] (lv4m) {dbp:Unitary\_state}
edge [->,black] node[text style]{qb4o:memberOf} (lv4)
edge [<-,black] node[text style]{sc:hasGovType} (lv1m);

\node[vertexr style=Gray, below=1.3cm of av1] (av3) {Unitary state}
edge [<-,black] node[text style]{sc:govName} (lv4m);

\end{tikzpicture}
\caption{\textsf{Citizenship} dimension: schema and sample instance.}
\label{fig:dimExample}
\end{figure*}

To conclude this section, Figure \ref{fig:dimExample} shows an excerpt of the representation of the \textsf{Citizenship} dimension schema using QB4OLAP. 
Again, empty circles represent blank nodes.
We also include a sample dimension instance on the right hand side of the figure. 
We can see that the property \texttt{qb4o:memberOf} is used to tell that Asia (\ttt{citDim:AS}) is  a member of the dimension level \textsf{Continent}. 
Note the relationship between schema and instance. 
For example, the property \ttt{sc:contName} 
is declared to be an attribute of the \textsf{Continent} level (\ttt{sc:continent}), 
and it is used to link a member of this level (Asia represented by the node \ttt{citDim:AS}), 
with the literal that represents its name. 
This example also shows how RUPs are defined in the schema and used in the instances. 
For example \ttt{sc:inContinent} is stated as the implementation of the RUP between the levels \textsf{Country} and  \textsf{Continent}, and it is used at the instance level to link members of these levels. 
\ref{appendixb} presents a complete QB4OLAP  representation of the Asylum applications data cube.

%With the inclusion of these concepts, QB4OLAP is capable of %representing the metadata needed to automatically implement 
%OLAP operations as SPARQL queries, as we explain later. 

\section{Querying QB4OLAP cubes}
\label{sec:queries}

We are now ready to get into the details of our 
approach for exploiting  data cubes on the SW, basically, 
enabling analytical queries. The rationale of our 
approach is based on the definition of a clear  separation 
between the conceptual and the logical 
levels, which, strangely, is not common in traditional OLAP.
On the contrary, popular OLAP query and analysis languages, 
such as MDX,  operate  at  the  logical  level  and  require, as 
we commented in Section \ref{sec:intro}, the user's deep 
understanding of how data are actually stored in order 
to be able to write queries. Further, even though MDX is a popular
language among OLAP experts, is far from being intuitive, 
and it would be a barrier for less technical users, who would like to 
manipulate a data cube to dive into the data. Thus, we follow an
approach \textit{aimed at promoting the data analysis directly on the SW},
and, for that, \textit{we want to allow analytical users to focus on querying 
QB4OLAP cubes} using the operations they know well, for example, roll-up 
or drill-down, to aggregate or dissagregate data, respectively,  
minimizing the need of dealing with technical aspects. 
Our hypothesis is that most users are hardly aware of SW models and
languages, but will easily capture the idea of languages dealing 
with cube operations. In addition, we consider, as explained, that
MDX is too technical for our ultimate goal explained above. 
Thus,  we  propose a high-level language, denoted \QL, based on an 
algebra for OLAP, whose only data type is the data cube.

\begin{figure}[!ht]
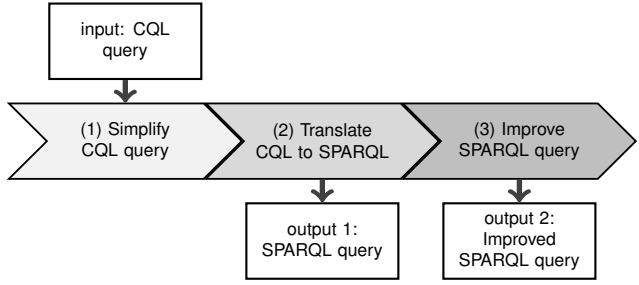


\begin{minipage}[c][4cm]{\linewidth}
\centering
\scriptsize
\smartdiagramset{
set color list={gray!10, gray!30,gray!50},
%  sequence item uniform color=gray!50,
  sequence item border color=black,
  sequence item text color=black,
  sequence item border size=1.2\pgflinewidth,
  sequence item font size=\scriptsize\sffamily,
  sequence item width=2cm,
additions={
additional item shape=rectangle,
%additional item fill color=gray!20,
additional item border color=black,
additional item offset=0.3cm,
additional item height=1cm,
additional arrow line width=2pt,
additional arrow tip=to,
additional arrow color=gray!50!black,
additional item font=\sffamily,
}
}
\smartdiagramadd[sequence diagram]{(1) Simplify CQL query,(2) Translate CQL to 
SPARQL,(3) Improve SPARQL query}
{above of sequence-item1/input: CQL query,below of sequence-item2/output 1: 
SPARQL query,below of sequence-item3/output 2: Improved SPARQL query}
\smartdiagramconnect{to-}{sequence-item1/additional-module1}
\smartdiagramconnect{-to}{sequence-item2/additional-module2}
\smartdiagramconnect{-to}{sequence-item3/additional-module3}
\end{minipage}
\caption{Query processing pipeline.} 
\label{figure.querypipeline}
\end{figure}

 Figure \ref{figure.querypipeline} shows the query processing pipeline.
The process starts with  a \QL~query that is first simplified (as explained in Section~\ref{sec:qlsimpl}). This stage 
aims at rewriting the query to eliminate unnecessary operations, or operations written in a sequence that is probably not the best one.\footnote{We remark that in a 
self-service BI environment~\cite{DBLP:journals/jdwm/AbelloDEGMNPRTVV13} users may not be experts, even to write queries in simple languages like \QL} 
 The second step translates the simplified \QL~query into a single 
 SPARQL expression, following a \textit{na{\"i}ve} approach (Section   
 \ref{sec:ql2sparql}). 
Finally, we apply SPARQL optimization heuristics to improve the 
performance of the \textit{na{\"i}ve} queries (Section \ref{sec:sparqlopt}).

\subsection{The \QL~language}
\label{sec:ql}
	 
\QL~follows the ideas introduced   by Ciferri et al.~\cite{CCG+13},
where a clear separation between the conceptual and the logical 
levels is made, allowing users to manipulate cubes regardless of
their underlying representation. In that paper, an algebra, denoted 
Cube Algebra, is sketched. \QL~is a subset of such algebra, and we chose it because it includes the most common OLAP operations.

We next  define a formal data model for cubes, and
define OLAP  operations in \QL~over this model. 
The model is based on the one proposed by Hurtado et al.~\cite{HGM2011}, 
although we choose a different way to present it, which allows 
to define the semantics of the operations in a clean and elegant way.
Due to space limitations, in the following we only present the main ideas to make this paper self-contained. We refer the reader to \cite{etcheverry2015} for   details.

%A cube expressed using this data model can be 
%represented using the QB4OLAP vocabulary, therefore we can 
%provide a SPARQL implementation
%of each \QL~operator. 

\begin{definition}(\textbf{Dimension schema}). A \textit{dimension 
		schema} is a tuple 
	$\langle  d, \mathcal{L}, \rightarrow, \mathcal{H}\rangle$ where:
	%\begin{enumerate}[label=(\alph*)]
	(a) $d$ is the name of the dimension; 
	(b) $\mathcal{L}$ is a set of pairs $\langle l, \mathsf{A}_l 
	\rangle$, called \textit{levels}, where $l$ identifies 
	a level in $\mathcal{L}$, and $\mathsf{A}_l= \langle a_1, \dots, a_n \rangle$ 
	is a tuple
	of level \textit{attributes}. Each attribute $a_i$ has a domain $Dom(a_i)$; 
	(c) `$\rightarrow$' is a partial order over the 
	levels in $\mathcal{L}$, with a unique bottom 
	level and a unique top level (\te{All}); 
	(d) $\mathcal{H}$ is a set of pairs $\langle h_n, L_h 
	\rangle$, called \textit{hierarchies}, where $h_n$ identifies the 
	hierarchy, $L_h$ is a set of levels such that  $L_h \subseteq \mathcal{L}$, and   
	there is at least one  
	path between the bottom level in $d$, and the top level  
	\te{All} composed of all the levels in $L_h$.\qed
	
	%\end{enumerate}
	%\vspace{-0.5cm}
	\label{def:dimSc}
\end{definition}

\begin{definition} (\textbf{Dimension instance}). Given a 
	dimension schema  $\langle d, \mathcal{L}, \rightarrow,  
	\mathcal{H} \rangle$, a \textit{dimension instance} $I_d$  
	is a tuple $\langle \langle d, \mathcal{L}, \rightarrow,  \mathcal{H} \rangle, \mathcal{T}_l, \mathcal{R} \rangle$  
	where: (a)   $\mathcal{T}_l$ is a  finite set of tuples of the form $\langle v_1, v_2, 
	\dots, v_{n} \rangle$, such that $\forall l$, $L=\langle l, \langle a_1, \dots, a_n \rangle  \rangle \in \mathcal{L},$
	and $\forall i, i=1,\dots,n, v_i \in Dom(a_i)$; 
	(b)   $\mathcal{R}$ is a finite set of relations, called  \emph{rollup}, 
	denoted $RUP^{L_j}_{L_i}, L_i,L_j \in L$,  
	where $L_i \rightarrow L_j~\in~$`$\rightarrow$',   
	\qed
	\label{def:dimIns}
\end{definition} 

\begin{definition} (\textbf{Cube schema}).
	Assume that there is a set $\mathcal{A}$ of aggregate functions
	(at this time we consider the typical SQL functions $\mbox{\sc{Sum}, \sc{Count}, \sc{Avg}, \sc{Max}, \sc{Min}},$ 
	%the ones addressed  in~\cite{Kuijpers2016}).  
	A \textit{cube schema} is a tuple 
	$\langle C_n, \mathcal{D}, \mathcal{M},\mathcal{F} \rangle$ where:
	%\begin{enumerate}[label=(\alph*)]
	(a)  $C_n$ is the name of the cube; 
	(b) $\mathcal{D}$ is a finite set of dimension schemas (cf. Def.~\autoref{def:dimSc}); 
	(c)  $\mathcal{M}$ is a finite set of attributes, where each   $m \in \mathcal{M}$, called \emph{measure}, has domain $Dom(m)$; 
	(d) $\mathcal{F}:\mathcal{M} \rightarrow \mathcal{A}$ is a function that maps measures  in $\mathcal{M}$ to an aggregate function 
	in $\mathcal{A}$. \qed
	\label{def:cubeSch}
\end{definition}

To define a cube instance we need to introduce the notion of \textit{cuboid}.  

\begin{definition} (\textbf{Cuboid instance}). Given: (a) A cube schema $\langle C_n, \mathcal{D},$  $\mathcal{M},\mathcal{F} \rangle$, where $|\mathcal{D}|= r$ and $|\mathcal{M}|= p$, (b) A dimension instance $I_{d_i}$ for each $d_i \in \mathcal{D}, i=1,\dots,r$; and (c) A set 
	of levels $\mathcal{V}_{Cb}=\lbrace L_1, L_2, \dots, L_D \rbrace$ 
	where $L_j \in \mathcal{L}_j$  in $d_i,  
	i=1,\dots,r$, such that   not two levels belong to the same 
	dimension, a \textit{cuboid instance}   is a partial function
	$Cb:\mathcal{T}_{L_1} \times \dots \times \mathcal{T}_{L_D} 
	\rightarrow Dom(m_1)\times \dots \times Dom(m_M)$, where $m_k \in \mathcal{M}, \forall k, k=1,\dots, p$. 
	The elements in the domain of $\mathsf{Cb}$ are called \textit{cells} (whose content are  elements in the range of $\mathsf{Cb}$), and 
	$\mathcal{V}_{Cb}$ is called  the \emph{level set} of the cuboid.
	\qed
	\label{def:cuboidInstance}
\end{definition}

We can now define a \textit{lattice of cuboids} referring  to the same cube schema, provided that we define an order between cuboids. We do this next.

\begin{definition} (\textbf{Adjacent Cuboids}). Two cuboids $\mathsf{Cb}_1$ and 
	$\mathsf{Cb}_2$, that refer to the same cube schema, 
	are \emph{adjacent} if their corresponding level sets $\mathcal{V}_{Cb_1}$ 
	and $\mathcal{V}_{Cb_2}$ differ in exactly one level, i.e., $|\mathcal{V}_{Cb_1} 
	-\mathcal{V}_{Cb_2}| = |\mathcal{V}_{Cb_2} -\mathcal{V}_{Cb_1}| = 1$. 
	\qed 
	\label{def:cuboidAdj}
\end{definition}

 \begin{definition} (\textbf{Order between adjacent cuboids}). \newline
	Given two adjacent cuboids 
	$\mathsf{Cb}_1$ and $\mathsf{Cb}_2$, such that $\mathcal{V}_{Cb_1} 
	-\mathcal{V}_{Cb_2} = \lbrace L_c \rbrace$ and $\mathcal{V}_{Cb_2} 
	-\mathcal{V}_{Cb_1} = \lbrace L_r \rbrace$, and $L_r$ and $L_c$ are levels in a dimension $d_k$ such that $L_c \rightarrow L_r$; then, we define the order  $\mathsf{Cb}_1 \preceq \mathsf{Cb}_2$ between both cuboids. 
	Moreover, for each pair of adjacent cuboids $\mathsf{Cb}_1 \preceq 
	\mathsf{Cb}_2$, each cell $c = (c_1, \dots, c_{k-1},c_k, c_{k+1}, \dots, c_n,$ 
	$ m_1,m_2, \dots m_p) \in \mathsf{Cb}_2$ 
	can be obtained from the cells in $\mathsf{Cb}_1$ as follows.
	Let $(c_1, \dots, c_{k-1},b_{k1},$ $c_{k+1}, \dots, c_n,$  $m_{1,1},m_{2,1}, \dots m_{p,1}), 
	(c_1,\dots,c_{k-1},b_{k2}, c_{k+1}, \dots,
	 c_n, m_{1,2},$ 
	 $m_{2,2},\dots,m_{p,2}), \dots, 
	(c_1,\dots,c_{k-1},b_{kq}, c_{k+1}, \dots, c_n, m_{1,q},$ 
	$m_{2,q},\dots m_{p,q})$  
	be all the cells in $\mathsf{Cb}_1$ where $(b_{k_i},c_k) \in RUP_{L_{c}}^{L{r}}, i=1\dots q$. Measures in  $c \in \mathsf{Cb}_2$ are 
	computed as $m_i = AGG_i(m_{i,1},\dots,m_{i,j}),$ $j=1..q$, where $AGG_i$ is the 
	aggregate function related to  $m_i$.
	\qed
	\label{def:cuboidOrder}
\end{definition}

A \textit{Cube Instance} is the lattice of all   
cuboids that share the same cube schema, defined over the
 $\preceq$ order relation above. The bottom of this lattice is the original cube, and the top  is the cuboid with 
just the \te{All} level for all the dimensions in the cube. If  $\mathsf{Cb}_i$ and $\mathsf{Cb}_j$ are two cuboids in the lattice, such that there is a path from $\mathsf{Cb}_i$ to
$\mathsf{Cb}_j$, we say that $\mathsf{Cb}_i \preceq^* \mathsf{Cb}_j$.  

Now, we are ready to give a precise   semantics for the operations in the OLAP algebra that will be the basis for ~\QL~(see~\cite{etcheverry2015} for details). 
 
The \textsc{Roll-up} operation  summarizes data to a higher level along a dimension hierarchy; that is, it receives a cuboid $\mathsf{Cb}_1$ in a cube instance, and a level $L$ in a dimension $D$, and returns another cuboid $\mathsf{Cb}_2$ in the same instance, such that $\mathsf{Cb}_1 \preceq^* \mathsf{Cb}_2$,  $L \in \mathcal{V}_{Cb_2}$, and 
$\mathcal{V}_{Cb_2} - \mathcal{V}_{Cb_1}= \{L\}.$
The \textsc{Drill-down} operation does the inverse, i.e., it receives a cuboid  $\mathsf{Cb}_1$, and a level $L$ in a dimension $D$, and returns a cuboid $\mathsf{Cb}_2$ such that 
$\mathsf{Cb}_2 \preceq^* \mathsf{Cb}_1$, and 
$\mathcal{V}_{Cb_2} - \mathcal{V}_{Cb_1}= \{L\}.$  
Note that  the cuboids resulting from a \textsc{Roll-up} or a \textsc{Drill-down} on a dimension $D$ are always reachable from
the bottom of the cube instance. Thus,  a \textsc{Drill-down} over  a dimension $D$ to a level $L$ can be obtained performing a \textsc{Roll-up} over $d$ from the bottom cuboid up to $L$.  
Since \textsc{Roll-up} and \textsc{Drill-down} only imply a navigation across a lattice (and do not modify it), 
we call them Instance Preserving Operations (IPO).

The \textsc{Dice} operation  selects the cells in a cube 
that 
satisfy a boolean condition $\phi$. It is analogous  
to the selection operation in the relational algebra. The  
condition  $\phi$ is expressed   over level member attributes, and/or  measure values.

The  \textsc{Slice} operation   removes 
one of the dimensions or measures in the cube. It is analogous to  the projection  
operation in relational algebra. In the case of eliminating a dimension, it is required that, before slicing, the dimension contains a single element at the instance level
~\cite{Agra97}. If this condition is not satisfied, 
a \textsc{Roll-up} to the \textsf{All} level   must be applied over this dimension before removing it.

We denote  operations \textsc{Dice} and \textsc{Slice} as  Instance Generating Operations (IGO), since they induce a new lattice (because they reduce the number of  cells in the cuboid, or  reduce the dimensionality of the cube, respectively), whose bottom cuboid is the result of the corresponding operation. Again, see~\cite{etcheverry2015} for   details.

 In the remainder, we will make use of the following properties. For the sake of space,
 we omit the proofs. 

\begin{property} (\textsc{Roll-up}/\textsc{Drill-down} commutativity)
A sequence of two  consecutive \textsc{Roll-up} (\textsc{Drill-down}) operations over 
different dimensions is commutative. \qed
\label{prop:ipocomm}
\end{property}

\begin{property} (\textsc{Roll-up}/\textsc{Drill-down} composition)
A sequence  of consecutive \textsc{Roll-up} and 
\textsc{Drill-down} operations over the same  dimension D, is equivalent to a   
\textsc{Roll-up} from the bottom level of   D, to  
the  level reached by the last operation  in  the sequence. \qed
\label{prop:ipocomp}
\end{property}

\begin{property} (\textsc{Roll-up}/\textsc{Drill-down} identity)
The application of the \textsc{Roll-up} or \textsc{Drill-down} operation over a 
dimension D from a level L to itself is equivalent to not applying the operation 
at all. \qed
\label{prop:ipoident}
\end{property}

\begin{property} (Slicing \textsc{Roll-up} and \textsc{Drill-down})
Performing a \textsc{Slice} operation over a dimension D after a sequence of 
\textsc{Roll-up} and \textsc{Drill-down} operations over D, 
is equivalent to apply only the \textsc{Slice} operation. \qed
\label{prop:iposl}
\end{property} 

A \QL~query~ is a sequence of  OLAP operations defined above, where the input cuboid 
of an operation is the output cuboid produced by the previous one.  
We assume that the input cuboid for the first operation  in the sequence    is  the bottom cuboid of a certain cube instance.  

\subsubsection{\QL~by example}
\label{sec:ql_notation}

We now present the syntax of a \QL~expression by means of an example.
Consider Query 1 below.

\begin{querynonum}
\label{ex:q32_secq}
\textbf{Query 1:} Total asylum applications submitted by African citizens to 
France in 2013, 
(by sex, time, age, and citizenship country)
\end{querynonum}

\begin{example} (\QL~query)
\label{ex:qlnotation}
The following \QL~query produces a cuboid that answers Query 1. 
For clarity, intermediate results are stored in variables $C_i$, 
although this is not mandatory. 
\begin{lstlisting}[style=qlquery]
$C1:=ROLLUP(migr_asyapp, timeDim, year);
$C2:=ROLLUP($C1,citizenshipDim,continent);
$C3:=DICE($C2,(citizenshipDim|continent|contName = "Africa"));
$C4:=DICE($C3,(destinationDim|geo|counName = "France" AND 
               timeDim|year|yearNum = 2013));
$C5:=DRILLDOWN($C4,citizenshipDim,citizenship);
$C6:=SLICE($C5,asylappDim);
$C7:=SLICE($C6,destinationDim);
\end{lstlisting}

First, a \textsc{Roll-up} operation  
aggregates measures up to the \te{Year} level in the \te{Time} dimension.
To keep only the cells that correspond to African citizens, a  \textsc{Roll-up} 
is performed over the \te{Citizenship} dimension, up to the 
\te{Continent} level;  
then a \textsc{Dice} operation  keeps cells corresponding to members of 
this level, that
satisfy the condition over the \te{contName} attribute. 
Another \textsc{Dice} operator  restricts
the results to cells that correspond to
France and to the year 2013. Then, a \textsc{Drill-down}   is applied 
to go back to the \te{Citizenship} level (the applicant's country).  
 Finally,  dimensions
\te{Application Type} and \te{Destination} are sliced out since we do not want them in the result. We remark that the user only deals with the elements of the MD model (e.g., cubes, dimensions), and not  the unfriendly (for non-experts) technical issues concerning MDX, SPARQL, RDF, etc. 
Also note the use of the notation dimension$|$level$|$attribute in the  
\textsc{Dice} expressions. \qed
\label{ex:query1} 
\end{example}

\subsubsection{Well-formed \QL~queries}
\label{sec:validql}

We define \textit{well-formed} \QL~queries as follows.

\begin{definition}(\textbf{Well-formed \QL~query}). A \textit{well-formed} \QL~query   satisfies the following conditions: (i) There is at most one \textsc{Slice} operation over each dimension D or measure M; (ii) Every \textsc{Drill-down} operation over a dimension D is preceded by at least one \textsc{Roll-up}  over the same dimension; (iii) There is no \textsc{Dice} operation mentioning  conditions over
measure values, in-between a  \textsc{Roll-up} and/or a  \textsc{Drill-down}. \qed
\label{def:wellformed}
\end{definition}

The reason why we prevent \textsc{Dice} operations including conditions over
measure values in-between a  \textsc{Roll-up} and/or \textsc{Drill-down}, is 
that we want to avoid 
storing additional information, in particular the computation trace. 
We illustrate this situation with the following example.

\begin{example} (Condition (iii) in Definition~\ref{def:wellformed})
\label{ex:qlint}
Consider the    query:
\begin{querynonum}
	\label{ex:q4}
	\textbf{Query 2:} Total asylum applications per month by sex, time, age, 
	citizenship, destination, and application type, only for years where the total 
	amount of applications is less than 100.
\end{querynonum}

The \QL~program below produces the answer to Query 2, although it is not 
well-formed. We next explain why.  
\begin{lstlisting}[style=qlquery]
$C1:=ROLLUP(migr_asyapp, timeDim, year);
$C2:=DICE($C1, obsValue < 100);
$C3:=DRILLDOWN($C2,timeDim, month);
\end{lstlisting}

First, a \textsc{Roll-up}  
aggregates measures up to the \te{Year} level on the \te{Time} 
dimension. Thus, the measure now contains the \textit{aggregated} values,
not the original ones. A \textsc{Dice} operation is then
applied to keep cells that satisfy
the restriction over the aggregated measure value. However, since
we want the results at 
the \te{Month} level, we  would need to keep 
track of the cells in the cuboid at the \te{Month} level, that roll up to 
the years that satisfy the \textsc{Dice} condition at the \te{Year} level. 
Condition (iii) in 
Definition~\ref{def:wellformed} prevents this.
\qed
\label{ex:query2} 
\end{example}

To summarize, the following patterns define the valid \QL~queries, using   regular expression notation.  
${\textsc{Dice}}_{l}$  and 
${\textsc{Dice}}_{m}$ denote \textsc{Dice} operations applied only over  level attribute  or measure values, respectively.

\begin{description}
\label{desc.qpatterns}
 \item \textbf{P1:} ${(\textsc{Slice}^*|\textsc{Dice}^*|\textsc{Roll-up}^*)}^+$
 \item \textbf{P2:} 
${({\textsc{Slice}}^*|{\textsc{Roll-up}}^+|{\textsc{Drill-down}}^+|{\textsc{
Dice}}_l^+)}^+$
 \item \textbf{P3:} 
${({\textsc{Slice}}^*|{\textsc{Roll-up}}^+|{\textsc{Drill-down}}^+|{\textsc{Dice
}}_{l}^*)}^+{\textsc{
Dice}}_{m}^+$
\end{description}

\subsection{\QL~simplification process}
\label{sec:qlsimpl}
As we have already  mentioned, \QL~is aimed at being used by non-experts. Thus, even well-formed \QL~queries may include unnecessary operations that should be eliminated. Further, operations can be reordered to reduce the size of the cuboid as early as possible. Based on the properties defined in Section \ref{sec:ql}, we define the following set of rewriting rules.  Between brackets we indicate the properties in which 
the rules are founded.

\textbf{Rule 1}. Remove all the \textsc{Roll-up} or \textsc{Drill-down} 
operations with the same starting and target levels (Property 
\ref{prop:ipoident}). 

\textbf{Rule 2}. Find sequences of \textsc{Roll-up} and/or \textsc{Drill-down} 
operations over 
the same 
dimension $D$, with no $\textsc{Dice}_{l}$ operation in-between, where 
$l$ is a level in $D$. Find the last level $l_D$ in the sequence. If $l_D$  is not the bottom level of D (call this
level ${l_b}_D$),   replace 
the sequence  with a single \textsc{Roll-up} from ${l_b}_D$ to $l_D$. 
Otherwise, remove all the operations in the group (Properties  
 \ref{prop:ipocomm} and \ref{prop:ipocomp}).

\textbf{Rule 3}. If there is  a \textsc{Slice} operation over a dimension $D$, and no \textsc{Dice} operation that mentions level members of  
$D$, move the \textsc{Slice} operation to the beginning
of the query; otherwise move it to the end.

\textbf{Rule 4}. If there is a \textsc{Slice} operation over a measure $M$, and 
  no \textsc{Dice} operation that mentions   $M$,  move the 
 \textsc{Slice}   to the beginning
of the query; otherwise move it to the end.

\textbf{Rule 5}. If there is a \textsc{Slice} operation over a dimension $D$,  
a sequence of \textsc{Roll-up} and \textsc{Drill-down} operations over 
  $D$, and 
  no \textsc{Dice} operation that mentions  levels   of  
  $D$,   remove all the \textsc{Roll-up} and 
  \textsc{Drill-down} operations, 
  and keep only the \textsc{Slice} operation (Property   \ref{prop:iposl}).

Let $q_{in}$ and $q_{out}$ be the \QL~query before and after the simplification process, respectively.
Then,  $q_{out}$ satisfies the following properties (proofs omitted).

\begin{property}
	\label{prop:struc3}
	If there is no \textsc{Dice} operation in $q_{in}$, there is at most one \textsc{Roll-up},
	and no \textsc{Drill-down} operation, for each Dimension $d$ in $q_{out}$. 
\end{property}

\begin{property}
	\label{prop:struc4}
	\textsc{Slice} operations are either at the beginning or at the end of $q_{out}$, but not in the middle. 
\end{property}

We now present an example of the simplification process, were we apply the rules above.

\begin{example} (\QL~simplification)
\label{ex:qlsimpl}
\begin{querynonum}
\label{ex:q31_secq}
\textbf{Query 3:} Total asylum applications per year (by sex, time, age, destination, and application type)
\end{querynonum}

The following \QL~expression  answers Query 3. 
\begin{lstlisting}[style=qlquery]
$C1:=ROLLUP(migr_asyapp, timeDim, year);
$C2:=ROLLUP($C1,destinationDim,government);
$C3:=ROLLUP($C2,citizenshipDim,continent);
$C4:=DRILLDOWN($C3,destinationDim,country);
$C5:=SLICE($C4,citizenshipDim);
\end{lstlisting}

The application of Rule 2 to   \$C2 and \$C4 replaces them with a single \textsc{Roll-up} on dimension \textsf{Destination}, from level \textsf{Country} to   itself, so it can be removed,  according to Rule 1. By Rule 3, operation \$C5 can be moved to the beginning of the query. Finally,  by Rule 5, we can remove \$C3, 
as operation \$C5 performs a \textsc{Slice} over the same dimension. The result  of the process is: 
 \begin{lstlisting}[style=qlquery]
$C1:= SLICE (migr_asyapp,citizenshipDim);
$C2:= ROLLUP ($C1, timeDim, year);
\end{lstlisting}
\vspace{-0.5cm}
\qed
\end{example}

\subsection{\QL~to SPARQL translation}
\label{sec:ql2sparql}

The next step in the process is the translation of \QL~queries (which are expressed at the conceptual level), into SPARQL expressions over QB4OLAP cubes (expressed at the logical level).
Our translation algorithms produce an SPARQL implementation of the \QL~operators. For this, we use the QB4OLAP representation of the formal model defined in Section \ref{sec:ql}, 
and the semantics of the operators defined in terms of this formal model. Recall that a cube instance \textsf{CB} is the lattice of all possible cuboids that adhere to a cube schema, 
and $\preceq$ is the partial order between adjacent cuboids in \textsf{CB}.
Definitions~\ref{def:cuboidAdj} and \ref{def:cuboidOrder} provide a mechanism to compute the cells of adjacent cuboids.
Therefore, starting from the bottom cuboid in the lattice (the one composed of the bottom levels in each dimension), all the  cuboids that form the cube instance can be computed  incrementally. Thus, to compute the \textsc{Roll-up} operation over an input cuboid $\mathsf{CB_{in}}$, it 
suffices to start at $\mathsf{Cb_{in}}$, and navigate the cube 
lattice visiting adjacent cubes that differ only in 
the level associated to dimension $D$, until we reach a 
cuboid $\mathsf{Cb_{out}}$, that contains the 
desired level in dimension $D$ (note that this path is unique, by definition).
 
We do not materialize   intermediate results. Instead, we directly compute the target cuboid
via a SPARQL query that navigates the dimension hierarchies up to the desired level, aggregating measure values using the aggregate functions declared in the QB4OLAP schema.
Note that this is a direct implementation of 
Definition~\ref{def:cuboidAdj} using SPARQL over a data cube represented using QB4OLAP.
Due to space limitations we do not present the translation algorithms  (which 
can be found in \cite{etcheverry2015}),  
but we present the ideas behind the 
implementation of each \QL~operator using SPARQL 1.1, by means of an example.

Let us consider  Query 4 below, and the \QL~query that expresses  it.\\

\begin{querynonum}
\label{ex:q33_secq}
\textbf{Query 4:} Total asylum applications per year submitted by Asian citizens to France or United Kingdom, 
where applications count $>$ 5000 
(by sex, time, age, citizenship country, and destination country)
\end{querynonum}

\begin{lstlisting}[style=qlquery]
$C1:=ROLLUP(migr_asyapp, citizenshipDim,continent);
$C2:=ROLLUP($C1, timeDim, year);
$C3:=DICE($C2,(citizenshipDim|continent|contName="Asia"));
$C4:=DICE($C3,( obsValue > 5000 AND 
			(destinationDim|country|counName = "France") OR 
			(destinationDim|country|counName="United Kingdom")));  
\end{lstlisting}

\begin{example} (\QL~to SPARQL translation)
\label{ex:ql2sparql}
The   SPARQL query below, produced by our translation algorithms, implements Query 4.
It contains a subquery, where aggregated values are computed, and an outer query where the \ttt{FILTER} conditions that implement the \textsc{Dice} operations are applied.
\begin{lstlisting}[style=sparql]
SELECT ?plm1 ?plm2 ?lm3 ?lm4 ?lm5 ?lm6 ?ag1
WHERE {
  { SELECT ?plm1 ?plm2 ?lm3 ?lm4 ?lm5 ?lm6  
           (SUM(xsd:integer(?m1)) as ?ag1) 
    FROM loc-ins:migr_asyapp_clean 
    FROM loc-sch:migr_asyappQB4O13 
    WHERE { ?o a qb:Observation .
            ?o qb:dataSet data:migr_asyapp .
            ?o sdmxm:obsValue ?m1 .
            ?o pr:citizen ?lm1 .
            ?lm1 qb4o:memberOf pr:citizen .
            ?lm1 sc:inContinent ?plm1 .
            ?plm1 qb4o:memberOf sc:continent .
            ?o sdmxd:refPeriod ?lm2 .
            ?lm2 qb4o:memberOf sdmxd:refPeriod .
            ?lm2 sc:inYear ?plm2 .
            ?plm2 qb4o:memberOf sc:year .
            ?o pr:geo ?lm3 .
            ?o pr:sex ?lm4 .
            ?o pr:age ?lm5 .
            ?o pr:asyl_app ?lm6 .
            ?plm1 sc:contName ?plm11 .
            ?lm3 sc:counName ?lm31 .
            FILTER ( ?plm11 = "Asia"  &&
	             (?lm31 = "France" || 
	              ?lm31  = "United Kingdom" ))}
     GROUP BY ?plm1 ?plm2 ?lm3 ?lm4 ?lm5 ?lm6 
} FILTER ( ?ag1 > 5000) } 
\end{lstlisting}
Lines 10 through 13 implement the first \textsc{Rollup} ($C1$). Variable  ?lm1 will be  instantiated with each member of the \te{Country} level in the \te{Citizen} dimension hierarchy, 
 related to an observation ?o (lines 10 and 11). Then, we navigate the hierarchy up to the  level \te{Continent}, using the rollup property  \ttt{sc:inContinent} (lines 12 and 13). The variable ?plm1 will contain the continent corresponding to the country 
 that instantiates  ?lm1. It is  placed in the \ttt{SELECT} clause  of the inner query (line 3),   
 in the \ttt{GROUP BY} clause of the inner query (line 27), and   in the result of the   
 outer query (line 1).
 Analogously, the navigation that corresponds to the \textsc{Rollup} in C2 is performed in   
 lines 14 through  17.
 % where 
%?plm2 gets a member of the \te{Year} and is added to the \ttt{GROUP BY} clause %and to the results of the inner and outer query. 
 Lines 18 to 21 will instantiate the level members of the remaining dimensions in the cube, which are also added to the \ttt{GROUP BY} clause, and to the \ttt{SELECT} clause of the inner and outer query. Line 9 retrieves the value of the measure in each observation, and the SUM aggregate function   computes ?xg1 in line 4. The aggregated value is added to the result of the outer query (line 1). In this case, measure values are converted to integer before applying the SUM function due to format restrictions of Eurostat data.
Finally, to implement the \textsc{Dice} operation in statement C3, we need  to obtain the name of each continent (line 22) and then use a \ttt{FILTER} clause to keep only the cells that correspond to ``Asia'' (line 24). The \textsc{Dice} operation in statement $C4$ is split as follows: the restriction on country names is implemented adding lines 25 and 26 to the \ttt{FILTER} clause (country names are retrieved in line 23), while the restriction on the measure values must be performed \textit{after} the aggregation, and is implemented by the \ttt{FILTER} clause of the outer query (line 28).  \qed
\end{example}

\subsection{SPARQL queries improvement}
\label{sec:sparqlopt}

We have shown a \textit{na\"{i}ve} procedure to automatically produce  SPARQL queries that  implement \QL~queries over QB4OLAP. To improve the performance of such queries, 
we  adapted three existing techniques to 
the characteristics of MD data in general, and of the QB4OLAP representation, in particular.
 
First, we adapted to our setting the  heuristics proposed by Loizou et al.\cite{LoizouAG15}  to  improve the performance of SPARQL queries. We next indicate the heuristics, and how 
we use some of them.

\textbf{H1 - Minimize optional graph patterns.} This heuristic is based on the fact that
the introduction of \ttt{OPTIONAL} clauses leads to PSPACE-completeness of the SPARQL evaluation problem\cite{Perez2009}. Since the SPARQL queries we produce do not include the OPTIONAL operator, we do not use this rule. 

\textbf{H2 - Use named graphs to localize SPARQL graph patterns.} This heuristic
is based on the  correlation 
between the performance of a query and the number of triples it is evaluated against. We apply this heuristic as follows. We  organize QB4OLAP data into two named graphs, namely: (a) A \textit{schema} graph, which stores the schema and dimension members; (b) An \textit{instance} graph, which stores only observations. Normally, the 
size of the instance graph will be considerably bigger than the schema graph. With this organization we can ensure a bound on the number of graph patterns over the 
instance graph, which will be at most 2+$|$D$|$+$|$M$|$, where D is the set of dimensions, and M the set of measures. 

\textbf{H3 - Reduce intermediate results.} This heuristic proposes to reduce intermediate results, replacing connected triple patterns with path expressions. This kind of patterns do not occur in our queries, and therefore this heuristic cannot be applied. This is due to the fact that QB4OLAP proposes to use a different predicate to represent each RUP relationship between level members, instead of using, as in QB, a single predicate like \ttt{skos:narrower}. We give an 
example of this in ~\ref{appendixb}.

\textbf{H4 - Reduce the impact of cartesian products.} This only applies when rows in the result differ in at most  one value. In those cases, it is suggested to collapse sets of almost identical rows into a single one, and to use aggregate functions. Since in the result of an OLAP query, each row represents exactly one point in the space (i.e., there is no redundancy), this heuristic cannot be applied to our problem.

\textbf{H5 - Rewriting \ttt{FILTER} clauses.} Proposes to transform \ttt{FILTER} clauses with disjunction ($||$) of equality constraints, using either the \ttt{UNION} of patterns, or a \ttt{VALUES} expression.  In Example~\ref{ex:h5} we show these transformations. Since the reported results are not conclusive on which of these strategies leads to better performant queries, we decided to evaluate both of them (see Section \ref{sec:eval}).

\begin{example} (Rewriting \ttt{FILTER} clauses)
\label{ex:h5} The queries below show how \ttt{FILTER} clauses with disjunction of equality constraints can be replaced using \textbf{H5}.  
\begin{lstlisting}[style=sparql]
SELECT ?x
WHERE { 
	?x <predicate> ?y .
    FILTER (?y = value1 || ?y = value2)}
#rewriting FILTER using UNION      
SELECT ?x
WHERE { 
	{ ?x <predicate> value1 }
	UNION
	{ ?x <predicate> value2 }	}
#rewriting FILTER using VALUES      
SELECT ?x
WHERE { 
	?x <predicate> ?y .
	VALUES ?y (value1 value2)}               
\end{lstlisting} \qed
\end{example}

As our second strategy, we considered the recommendations in  \cite{Vesse2014}, namely: 
 (i) Split conjunctive \ttt{FILTER} equality constraints into a cascade of \ttt{FILTER} equality constraints; (ii) Replace a \ttt{FILTER} equality constraint that compares a variable and a constant, with a graph pattern. 
The first recommendation may help the query processor to push \ttt{FILTER} constraints down in the query tree, while the second one allows the query processor to use indexes to select the patterns that match the criteria.  
\begin{example} (Improving FILTERs) Below, we give an example of the second strategy.
\label{ex:hjena} 
\begin{lstlisting}[style=sparql]
SELECT ?x
WHERE { ?x ?y ?z .
	FILTER (?y = <predicate> && ?z > value1)}
#splitting FILTER conjunction      
SELECT ?x
WHERE {	?x ?y ?z .
	FILTER (?y = <predicate>)
	FILTER (?z > value1)}
#replace FILTER equality constraints  with a  BGP   
SELECT ?x
WHERE {	?x <predicate> ?z.
	FILTER (?z > value1)}               
\end{lstlisting}
The query in Lines 1 to 3  asks for the values of ?x that are associated
 via $<$predicate$>$, with values greater that `value1'. We then rewrite the query applying the strategies mentioned above, i.e.,  splitting and rewriting.\qed
\end{example}

The next example shows the result of applying the above two   strategies to the query in Example \ref{ex:ql2sparql}.

\begin{example} (SPARQL queries improvement)
\label{ex:sparqlimp}
The application of \textbf{H2} organizes graph patterns in the inner query in two   
 \ttt{GRAPH} clauses: one that corresponds to patterns in the instance graph (lines 8  
 to 15), and another in the
 schema graph (lines 16 to 26). Applying \textbf{H5},
 the \ttt{FILTER} clause on country names is replaced by a \ttt{VALUES} clause (line 25). Finally, using the second  strategy, \ttt{FILTER} clauses are split, and the one on  
 continent name is  replaced by a graph pattern (line 20). 
\begin{lstlisting}[style=sparql]
SELECT  ?plm1 ?plm2 ?lm3 ?lm4 ?lm5 ?lm6 ?xg1
WHERE {
  {SELECT ?plm1 ?plm2 ?lm3 ?lm4 ?lm5 ?lm6  
	(SUM(xsd:integer(?m1)) as ?xg1) 
  FROM NAMED loc-ins:migr_asyapp_clean
  FROM NAMED loc-sch:migr_asyappQB4O13 
  WHERE { 
    {GRAPH loc-ins:migr_asyapp_clean
      {?o a qb:Observation .
	   ?o qb:dataSet data:migr_asyapp .
	   ?o sdmxm:obsValue ?m1 .
	   ?o pr:citizen ?lm1 .
	   ?o sdmxd:refPeriod ?lm2 .
	   ?o pr:geo ?lm3 . ?o pr:sex ?lm4 .
	   ?o pr:age ?lm5 . ?o pr:asyl_app ?lm6 .}}.
    {GRAPH loc-sch:migr_asyappQB4O13 
      {?lm1 qb4o:memberOf pr:citizen .
	   ?lm1 sc:inContinent ?plm1 .
	   ?plm1 qb4o:memberOf sc:continent .
	   ?plm1 sc:contName "Asia" .
	   ?lm2 qb4o:memberOf sdmxd:refPeriod .
	   ?lm2 sc:inYear ?plm2 .
	   ?plm2 qb4o:memberOf sc:year .
	   ?lm3 sc:counName ?lm31 .
	   VALUES ?lm31 {"France"@en "United Kingdom"@en}
  }}}
  GROUP BY ?plm1 ?plm2 ?lm3 ?lm4 ?lm5 ?lm6 
  } FILTER (?xg1 > 5000) }
\end{lstlisting}
\qed
\end{example}
Our third, and final, strategy, is based  on the work of Stocker et. al \cite{Stocker2008}.  
%We   propose to reorder triple patterns on the schema graph to further improve the performance of  SPARQL queries.
This optimization is based on 
graph pattern selectivity. The idea behind this approach is to reduce intermediate results by first applying  the most selective patterns. This requires 
to keep estimates on the selectivity of each pattern. In our case,
 we   take advantage of MD data characteristics to estimate 
the selectivity of patterns beforehand: Since typically, RUP relationships
between level members are functions, each level member has exactly one parent on
 the level immediately above. Thus, for
 each pair of levels 
${L}_i$ and ${L}_j$ such that $L_i \rightarrow 
 L_j$ in a hierarchy H, $|L_i| \geq |L_j|$. Moreover, 
in most cases $|L_i| > |L_j|$ holds. 
 Based on the above, we define alternative \textit{ordering criteria} (OC) for the graph patterns.

\begin{itemize}[noitemsep,leftmargin=5pt]
\item \textbf{Ordering Criterion 1 (OC1)} - For each dimension  appearing in the query,  apply first the patterns that correspond to higher levels.
\item \textbf{Ordering Criterion 2 (OC2)} - For each dimension, apply OC1. Then, reorder dimensions as follows:  First consider dimensions with conditions that fix a certain member, then dimensions with conditions that restrain to a range of members, and then the other dimensions.  
\item \textbf{Ordering Criterion 3 (OC3)} - For each dimension apply OC1. Then, reorder dimensions according to OC2.
If more than one dimension satisfy any of the criteria  in OC2, then use the number of members in the highest level reached for each dimension to decide the relative order between these dimensions. For example: If dimension A and dimension B fix members $a$ and $b$ at levels $l_A$ and  $l_B$ respectively, and $|l_A| \geq |l_B|$, then dimension A goes before dimension B.
\end{itemize}
 
\begin{example} (Reordering triple patterns)
\label{ex:sparqlOC}
We show the result of applying OC2 to reorder the triple patterns on the schema graph from 
Example \ref{ex:sparqlimp}.
\begin{lstlisting}[style=sparql]
GRAPH loc-sch:migr_asyappQB4O13 {  
  ?plm1 sc:contName "Asia" .
  ?plm1 qb4o:memberOf sc:continent .
  ?lm1 sc:inContinent ?plm1 .
  ?lm1 qb4o:memberOf pr:citizen .
  ?lm3 sc:counName ?lm31 .
  VALUES ?lm31 {"France"@en "United Kingdom"@en}
  ?plm2 qb4o:memberOf sc:year .
  ?lm2 sc:inYear ?plm2 .
  ?lm2 qb4o:memberOf sdmxd:refPeriod .}
\end{lstlisting}
 Triples in lines 2 through 5 correspond to the \te{Citizenship} dimension, lines 6 and 7  correspond to \te{Destination} dimension, and lines 8 through 10 correspond to the  \te{Time} dimension. For each dimension, the graph patterns are ordered from higher levels in the hierarchy to lower ones. Then, the relative position of each dimension in the query is altered with respect to the naive query. The \te{Citizenship} dimension is considered first since a member of the dimension is fixed to ``Asia''. Then we consider the  \te{Destination} dimension because there is a restriction on members of this dimension (``France'' or ``United Kingdom''). \qed
\end{example}

We  end this section with some remarks on the complexity of the generated SPARQL queries.
It has been proved that the evaluation of a SPARQL 1.0 
query is NP-complete for the AND-FILTER-UNION fragment of the language\cite{Perez2009}. 
Moreover, the evaluation of queries that only contain AND and UNION operators is 
already NP-complete, as proved in \cite{Schmidt2010}. Perez et. al \cite{Perez2009} also proved 
that the main source of complexity in SPARQL 1.0 queries is the introduction of the OPTIONAL, which leads to PSPACE-completeness of the evaluation problem. 
The SPARQL queries we produce, both \textit{na\"{i}ve} and improved, avoid the OPTIONAL operator but make an intensive use of two functionalities incorporated in SPARQL 1.1: The computation of 
aggregates (GROUP BY clauses), and subqueries. To the best of our knowledge 
there are still no theoretical results on the complexity of such queries, and a study of this issue  is beyond the scope of this work.

\section{Implementation}
\label{sec:impl}

\begin{figure}[h]
\centering
\includegraphics[width=\linewidth]{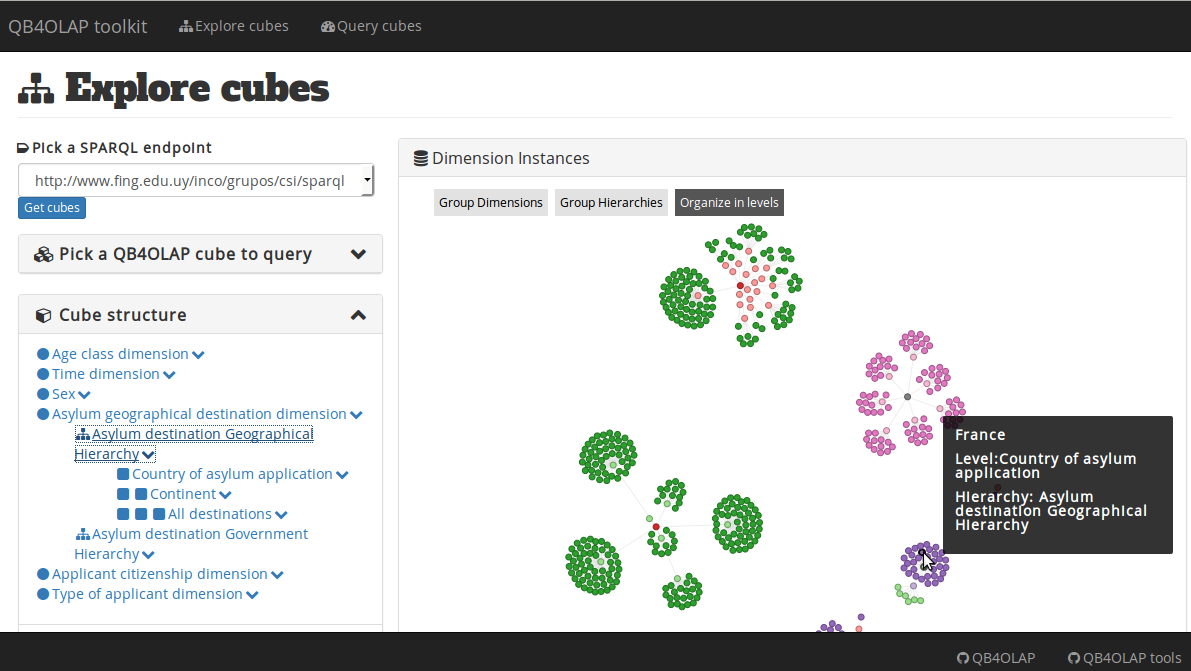}
\caption {QB4OLAP toolkit: Explorer module}
\label{fig:screenExpl}
\end{figure}

The \textit{QB4OLAP toolkit} is a web application  that implements our approach, allowing to explore and query QB4OLAP cubes. It is composed of two modules. 
The \textit{Explorer module}  enables the user to navigate the cube schema, 
and   visualize dimension instances stored in a SPARQL endpoint. 
Figure \ref{fig:screenExpl} presents a screenshot of this module. 

\begin{figure*}[t]
\centering
\includegraphics[width=\linewidth]{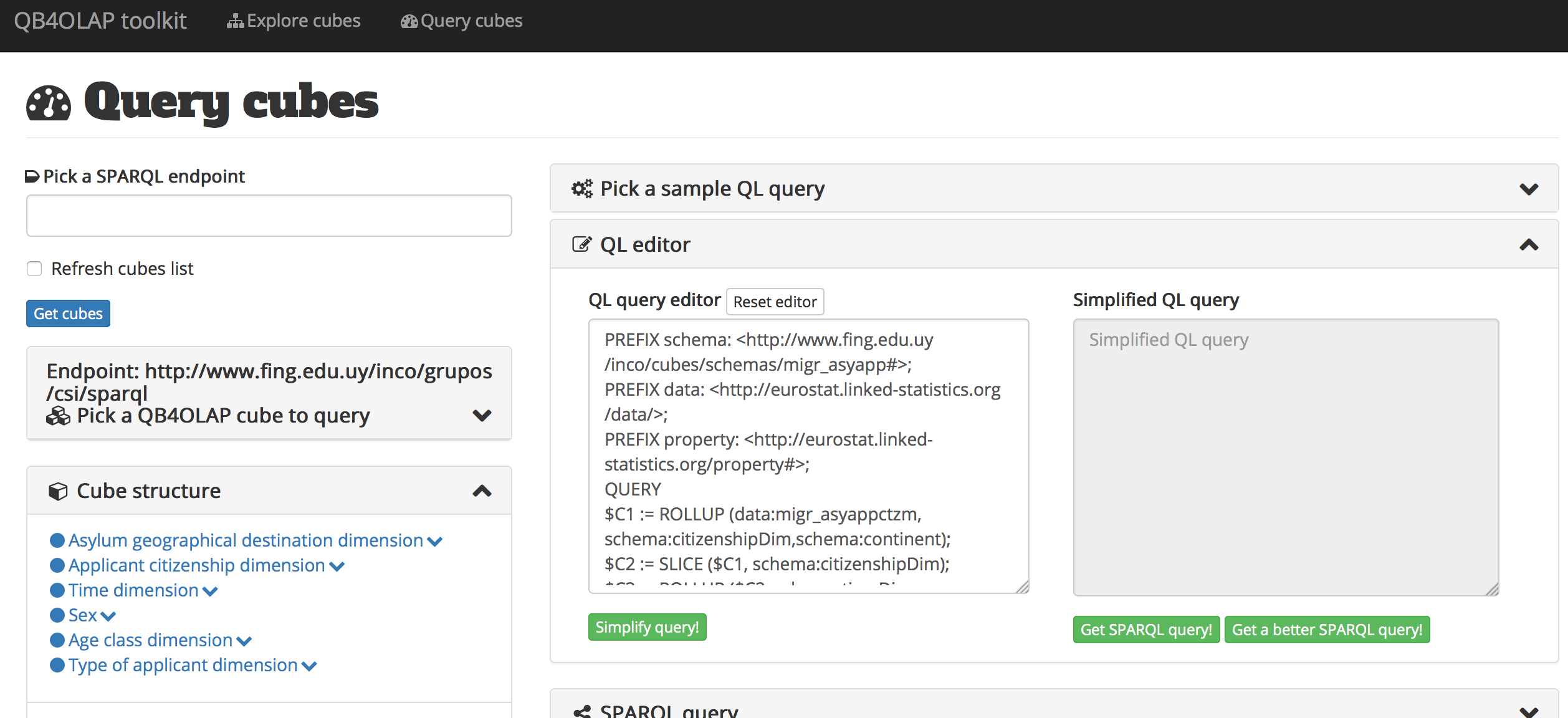}
\caption {QB4OLAP toolkit: Querying module}
\label{fig:screenquery}
\end{figure*}

The  \textit{Querying module} implements the querying processing pipeline presented in Figure~\ref{figure.querypipeline}. The user first writes  a \QL~query. Then, the
 application simplifies 
 this \QL~query, and displays  the result to the user, who  can 
 choose to generate either a na\"{i}ve SPARQL query or an improved  one. The 
 query produced is presented to the user and   executed.
 Results are presented in tabular format. 
 Figure \ref{fig:screenquery} presents a screenshot of this module. 
 
\begin{figure}[h]
\centering
\includegraphics[width=0.7\linewidth]{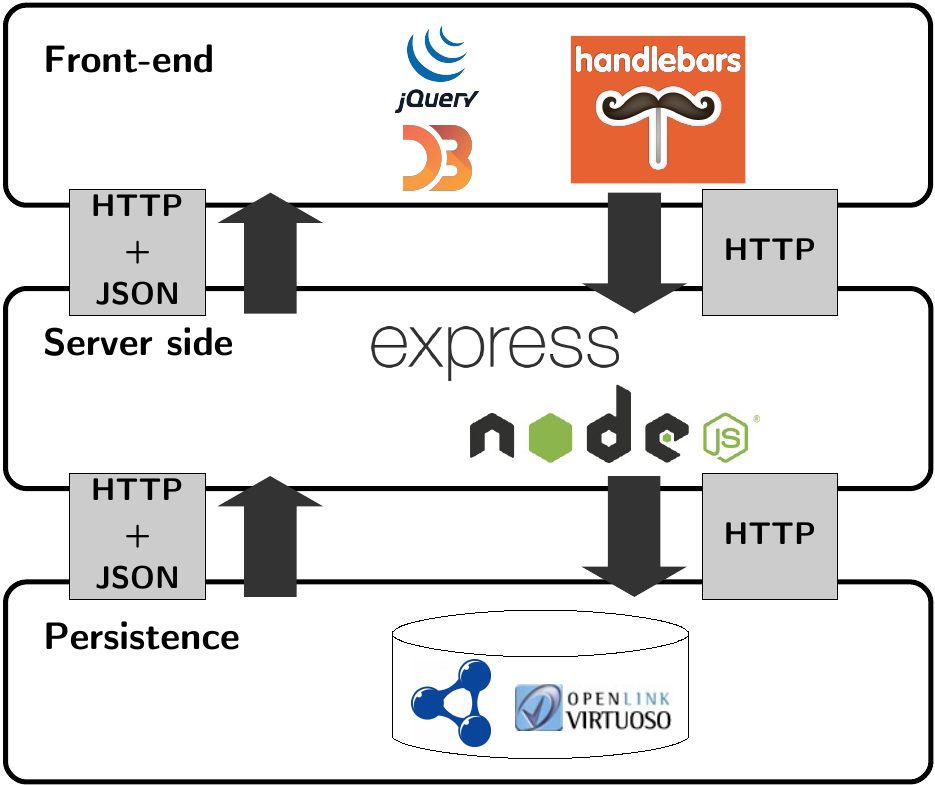}
\caption {QB4OLAP toolkit: technology stack}
\label{fig.archi}
\end{figure}

The QB4OLAP toolkit  has been entirely developed in Java Script over the Node.js platform using 
Express.  Handlebars, jQuery, and  D3.js are used to implement the front-end.  Virtuoso Open Source version 7 is used for RDF storage and SPARQL back-end. The 
communication with Virtuoso is implemented via HTTP and using JSON format to 
exchange data. Figure \ref{fig.archi} presents the technology stack of QB4OLAP 
toolkit.

The QB4OLAP toolkit is available online.\footnote{\url{https://www.fing.edu.uy/inco/grupos/csi/apps/qb4olap/}}   We also provide example queries that the user will edit and run. 
Source code is available at
GitHub.\footnote{\url{https://github.com/lorenae/qb4olap-tools}}

\section{Evaluation}
\label{sec:eval}

We now report and discuss experimental results. Our  primary goal is to show that, with our proposal, OLAP users can   write complex analytical queries in an algebra that is familiar to them,  manipulating just what they know well: data cubes, regardless of how they are physically stored. For what we are interested in this paper, OLAP users should be able to query cubes on the SW, without having to  deal with technical issues such as QB4OLAP, RDF, or SPARQL, and still obtain good query performance. 
%Note that \QL~programs just consist in sequences of OLAP operations, well known %to any OLAP user.
% Even further, this does not prevent the use of graphic tools that can translate %cube navigation paths to \QL~programs, thus providing independence of the %underlying data model.

Our evaluation goal is thus twofold: On the one hand, we want to compare our approach against other one(s) that are aimed at querying OLAP cubes on the web. On the other hand, we look for the best possible combinations of  query optimization strategies. For the first goal, we compare our approach against the one by K\"{a}mpgen et al.~\cite{Kampgen2013,Kampgen2012}, who propose a mechanism for implementing some OLAP operations over extended QB cubes using SPARQL queries (see Section \ref{sec:related} for details).
To evaluate their approach, they adapted the Star Schema Benchmark (SSB)\cite{Neil2009}, and produced the  SSB-QB benchmark, which 
consists of: (i) A representation of the SSB cube schema and dimension instances using QB and other related vocabularies;  
(ii) A representation of SSB facts as QB observations;  (iii) A  set of 
thirteen SPARQL queries over these data. These queries are equivalent to SSB queries, and aim at representing the most common types of star schema queries in an OLAP setting. Based on this work, we built the SSB-QB4OLAP benchmark, which consists of: 
(i) A representation of the SSB cube schema and dimension instances using QB4OLAP; (ii) The same observations as in SSB-QB;   (iii) 
 A  set of thirteen \textit{\QL~queries} that are \textit{equivalent to the  SSB-QB} queries (and also to the SSB queries). Thus, the 
SSB-QB4OLAP benchmark  allows us to  compare  our approach against~\cite{Kampgen2013}. It also allows us 
to measure the impact of our improvement strategies, in order 
 to address  our second goal.
For this, we   translated the   \QL~queries into SPARQL using the na\"{i}ve approach, and explore which combination of strategies yields the best  query results, based on several metrics. 

Next, we  introduce the SSB-QB4OLAP Benchmark (Section \ref{sec:eval.bench}),   describe the experimental setup 
and  experiments (Section \ref{sec:eval.setup}), 
and  discuss  the  results (Section \ref{sec:eval.discussion}).
The complete experimental environment is available for download as a virtual machine at the the benchmark 
site.\footnote{\url{https://github.com/lorenae/ssb-qb4olap}}

\subsection{The SSB-QB4OLAP Benchmark}
\label{sec:eval.bench}
 
\begin{figure}[t]
\centering
\includegraphics[width=0.8\linewidth]{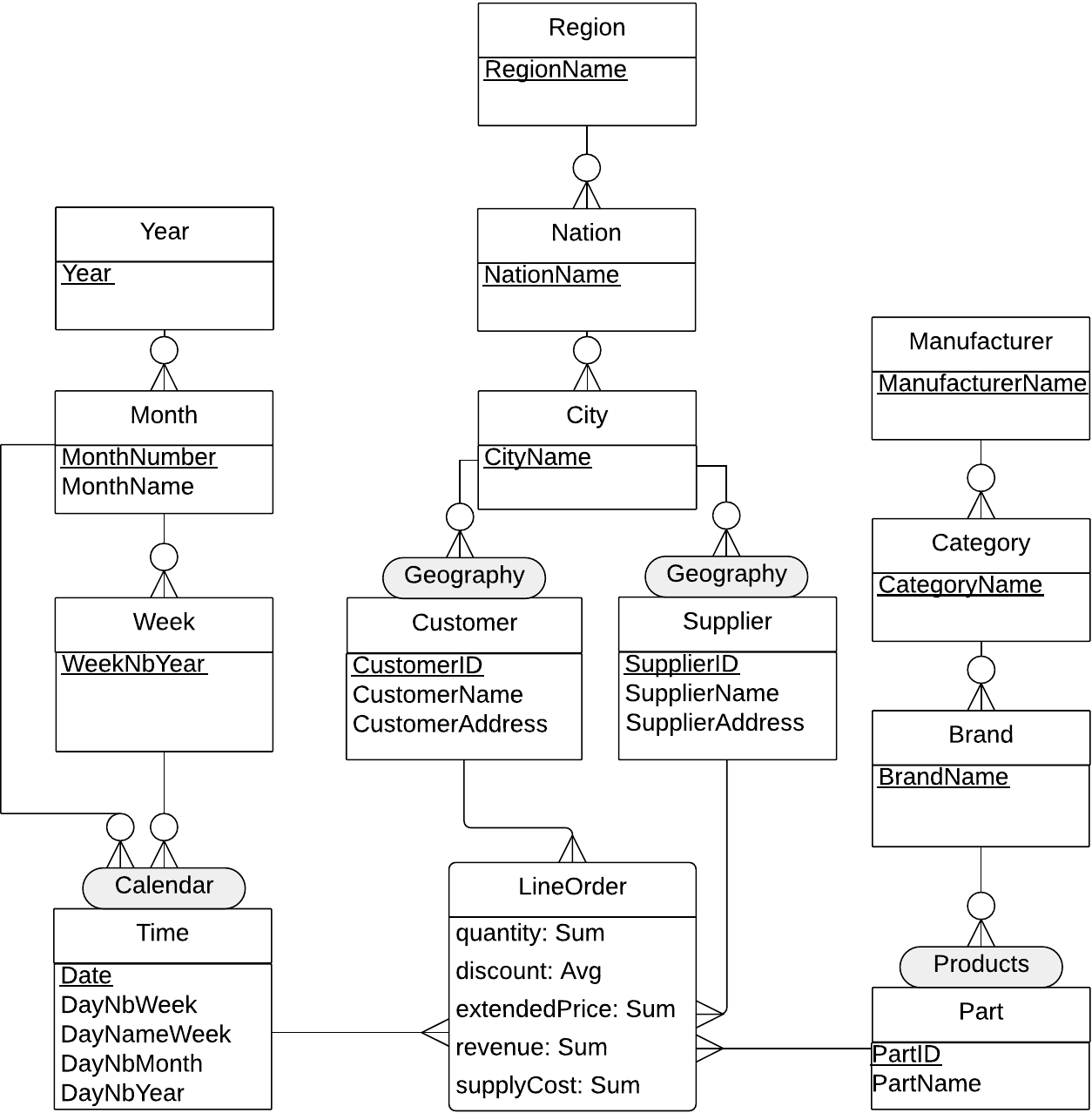}
\caption {Conceptual schema of the SSB-QB4OLAP cube}
\label{fig.concSSB}
\end{figure}

\textbf{SSB-QB4OLAP data~}
SSB-QB4OLAP represents the SSB data cube at Scale 1, and is organized in three sets of triples that represent: 
(1) Facts (observations); (2) The cube schema; and (3) The dimension instances (i.e., 
level members, attribute values, and RUP relationships). 
The set of observations,  as in SSB-QB,  consists of about  132,000,000 triples,  representing 6,000,000 line orders.
The cube schema is represented in QB4OLAP, consists of about  250 triples, and corresponds to the conceptual schema presented in Figure \ref{fig.concSSB}.
Each line order contains five measures (\textsf{quantity}, 
\textsf{discount}, \textsf{extended price}, \textsf{revenue}, and \textsf{supply 
cost}), which can be analyzed along four dimensions: \textsf{Time}, 
\textsf{Part}, \textsf{Customer}, and \textsf{Supplier}. 
Finally, a set of about  2,800,000 triples represents level members, attribute values, and 
rollup relationships. Table \ref{tab.ssbdata} shows the number of members in each 
level. 
Data are  available for querying at our endpoint.\footnote{\url{https://www.fing.edu.uy/inco/grupos/csi/sparql}}

\begin{table}[h]
\caption{SSB-QB4OLAP dataset statistics}
\label{tab.ssbdata}
\centering
\scriptsize
\begin{tabularx}{\linewidth}{|l|X|r|l|X|r|}
\hline
Dim.&Level&\#members&Dim.&Level&\#members\\
\hline
\multirow{4}{*}{Time}&Time&2556&\multirow{4}{*}{Part}&Part&2000000\\ 
\cline{2-3} \cline{5-6}
& Week & 371&&Brand&1000\\ \cline{2-3} \cline{5-6}
& Month & 84&&Cat.&25\\ \cline{2-3} \cline{5-6}
& Year & 7&&Manuf.&5\\ \hline
\multirow{4}{*}{Custom.}&Custom.&30000&\multirow{4}{*}{Supp.}&Supplier&2000\\ 
\cline{2-3} \cline{5-6}
& City & 250&& City & 250\\ \cline{2-3} \cline{5-6}
& Nation & 25&& Nation & 25\\ \cline{2-3} \cline{5-6}
& Region & 5&& Region & 5\\ \hline
\end{tabularx}

\end{table}

\textbf{SSB-QB4OLAP queries~} Queries are organized in four so-called  \textit{query 
flights}, which represent  different  types of usual star schema queries  (functional coverage), and to access varying fractions of the set of line orders 
(selectivity coverage). The \textbf{first query flight (QF1)} is composed of  three queries 
($Q_1$-$Q_3$) that impose restrictions on only one dimension, and quantify the revenue increase that would have resulted from eliminating certain 
company-wide discounts in a range of products in a certain year.
The three queries in the \textbf{second query flight (QF2)} 
($Q_4$-$Q_6$)   impose restrictions on two dimensions, and compare revenue for some 
product classes, for suppliers in a certain region, grouped by more restrictive 
product classes, along all years. The \textbf{third query flight (QF3)} 
has four queries ($Q_7$-$Q_{10}$) that impose restrictions on three dimensions, and aims at 
  providing revenue volume for line order transactions by customer 
nation, supplier nation, and year within a given region, in a certain time 
period. The \textbf{fourth query flight (QF4)} has three queries 
($Q_{11}$-$Q_{13}$) and restrictions over four dimensions. It represents a ``what if'' 
sequence of operations  analyzing  the profit for customers and suppliers 
from America on specific product classes over all years.

\subsection{Experimental setup and results}
\label{sec:eval.setup}

We ran our evaluation on an Ubuntu Server 14.04.1 LTS, a single Intel(R) Xeon(R) E5620  @2.40GHz with 4 cores and 8 hardware threads, 
32GB RAM, and 500GB for local data storage. We use Virtuoso Open source (V 07.20.3214) as RDF store.
BIBM tool\footnote{\url{http://sourceforge.net/projects/bibm/}} was used to perform \textit{TPC-H power tests}, and in each
test suit, a  mix of 13 queries was used with scale 1 and 2 client streams. We also ran  a test suit using the query mix from SSB-QB. We  measured
the average response time for each query and the following TPC-H metrics for each query mix:
\textit{TPC-H Power}, which measures the query processing power in queries per hour (QphH); 
\textit{TPC-H Throughput} (QphH),   the total 
number of queries executed over the length of the measurement interval; and \textit{TPC-H Composite},  the geometric mean of the previous metrics, that reflects the query processing power when queries are submitted in a single stream, and  the query throughput for queries  submitted by multiple concurrent users~\cite{TPCH}. 

%In Section \ref{sec:eval.setup.imp} we  report   
%the experiments performed to evaluate the SPARQL improvement strategies,
%and Section \ref{sec:eval.setup.comp}  compares our approach with SSB-QB.

\subsubsection{Evaluation of the improvement strategies}
\label{sec:eval.setup.imp}

We measured the impact on performance of the improvement strategies presented in Section \ref{sec:sparqlopt}, in order to find out which combination of strategies results more beneficial. The 
strategies are summarized  in Table \ref{tab.spHeuristics}. 

\begin{table}[h]
\caption{Strategies used to improve queries performance}
\label{tab.spHeuristics}
\centering
\scriptsize
\begin{tabularx}{\linewidth}{|X|}
\hline
S1:~Use named graphs to reduce the search 
space~\cite{LoizouAG15}\\\hline
S2:~Replace FILTER equality constraints that compare a variable and a constant with BGPs~\cite{Vesse2014}\\\hline
S3:~Split FILTER clauses with CONJUNCTION of constraints into a cascade of FILTER clauses with atomic
 constraints~\cite{Vesse2014}\\\hline
S4:~Replace FILTER clauses with DISJUNCTION of equality constraints using UNION or VALUES~\cite{LoizouAG15}\\\hline
S5:~Reorder triple patterns applying most restrictive patterns for each dimension first (using criteria OC1, OC2, or OC3)\\\hline
\end{tabularx}

\end{table}

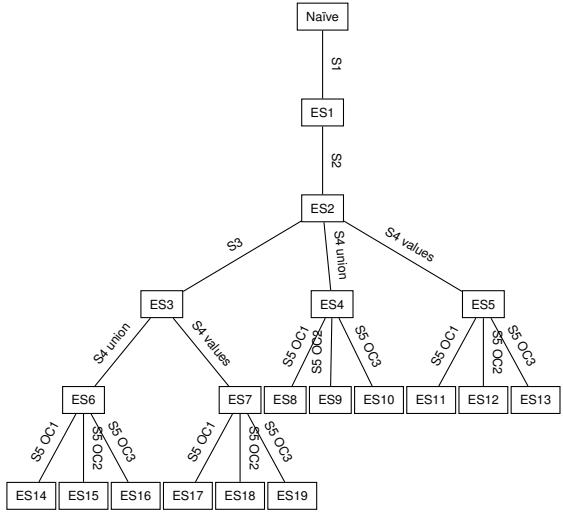
\begin{figure}[t]
\centering
\begin{forest}
for tree={
  font=\sffamily\tiny,
    l sep=.9cm,
    s sep=0.03cm,
    delay={
      edge label/.wrap value={node[midway,above,sloped,font=\sffamily\tiny, above]{#1}},
    },
    %minimum height=0.2cm,
    %maximum width=1cm,
    draw %Put lines around each
},
[Na\"{i}ve
  [ES1,edge label=S1
    [ES2,edge label=S2
       [ES3 ,edge label=S3
         [ES6 ,edge label={S4 union}
           [ES14 ,edge label={S5 OC1}] 
           [ES15 ,edge label={S5 OC2}] 
           [ES16 ,edge label={S5 OC3}] 
         ]
         [ES7 ,edge label={S4 values}
           [ES17 ,edge label={S5 OC1}] 
           [ES18 ,edge label={S5 OC2}] 
           [ES19 ,edge label={S5 OC3}] 
         ]
       ]
       [ES4 ,edge label={S4 union}
        [ES8 ,edge label={S5 OC1}] 
        [ES9 ,edge label={S5 OC2}] 
        [ES10 ,edge label={S5 OC3}] 
       ]
       [ES5 ,edge label={S4 values}
        [ES11 ,edge label={S5 OC1}] 
        [ES12 ,edge label={S5 OC2}] 
        [ES13 ,edge label={S5 OC3}] 
       ]
     ]
   ]  
]
\end{forest}
\caption{Improvement Strategies Evaluation Scenarios}
\label{fig.evalscenarios}
\end{figure}

For each of the 13 queries in the benchmark, Table \ref{tab.imp_app} indicates which strategies in  Table~\ref{tab.spHeuristics} can be applied to them. The combination of all possible strategies defines a space from which we chose a subset, based on the applicability of the strategies to the different queries. Thus, we devised a space of \textit{Evaluation Scenarios (ES)}, where each scenario represents the application of a sequence of improvement strategies to the na\"{i}ve SPARQL queries. 
Figure \ref{fig.evalscenarios} shows the space of evaluation scenarios as a tree. Each node represents an ES, and labels on edges represent the improvement strategy applied to transform a parent ES into a child ES. 
We can see that S1 and S2 were chosen to belong to all evaluation scenarios, since they apply to most queries (as we can see in Table \ref{tab.imp_app}). Then we consider the cases of applying S3 (ES3) or not. For S4 we consider both flavours: either replacing FILTER conjunction with UNION or VALUES clauses. 
Finally, we consider the triples reordering strategy (S5) using each of the ordering criteria discussed in Section \ref{sec:sparqlopt}. 
As an example, ES11 is the result of applying improvement strategies S1, S2, S4 (VALUES) and S5 (OC1), to na\"{i}ve SPARQL queries. 

\begin{figure}[t]
  \begin{center}
  \footnotesize
    \fontsize{6}{6}\selectfont
     \includegraphics[width=\linewidth]{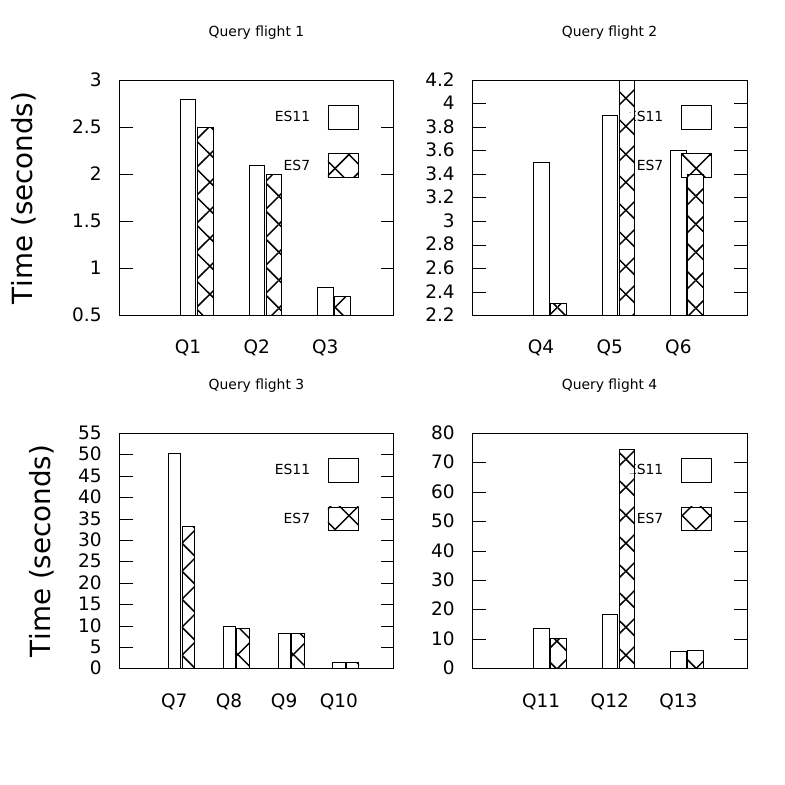}
     \vspace{-1cm}
    \caption{Na\"{i}ve  vs. improved queries execution time}
    \label{graph:Naive_B}
  \end{center}
\end{figure}

Table \ref{tab.tpcres} reports the results  for the
 na\"{i}ve approach and all the evaluation scenarios.   
 ES7 and ES11 are the scenarios with better performance. Figure \ref{graph:Naive_B} reports the average execution time for each query at the best improvement scenarios.

\begin{table}[h]
	\scriptsize 
	\begin{tabular}{L{0.06cm}|L{0.06cm}|L{0.06cm}|L{0.06cm}|L{0.06cm}|L{0.06cm}|L{0.06cm}|L{0.06cm}|L{0.06cm}|L{0.06cm}|L{0.18cm}|L{0.18cm}|L{0.18cm}|L{0.18cm}|}

		\cline{2-14}
		&$Q_1$&$Q_2$&$Q_3$&$Q_4$&$Q_5$&$Q_6$&$Q_7$&$Q_8$&$Q_9$&$Q_{10}$&$Q_{11}$&$Q_{12}$&$Q_{13}$\\\hline
		\multicolumn{1}{|c|}{S1}&\cellcolor{SeaGreen!50}$\checkmark$&\cellcolor{SeaGreen!50}$\checkmark$&\cellcolor{SeaGreen!50}$\checkmark$&\cellcolor{SeaGreen!50}$\checkmark$&\cellcolor{SeaGreen!50}$\checkmark$&\cellcolor{SeaGreen!50}$\checkmark$&\cellcolor{SeaGreen!50}$\checkmark$&\cellcolor{SeaGreen!50}$\checkmark$&\cellcolor{SeaGreen!50}$\checkmark$&\cellcolor{SeaGreen!50}$\checkmark$&\cellcolor{SeaGreen!50}$\checkmark$&\cellcolor{SeaGreen!50}$\checkmark$&\cellcolor{SeaGreen!50}$\checkmark$\\\hline
		\multicolumn{1}{ |c|}{S2}&\cellcolor{SeaGreen!50}$\checkmark$&\cellcolor{SeaGreen!50}$\checkmark$&\cellcolor{SeaGreen!50}$\checkmark$&\cellcolor{SeaGreen!50}$\checkmark$&\cellcolor{SeaGreen!50}$\checkmark$&\cellcolor{SeaGreen!50}$\checkmark$&\cellcolor{SeaGreen!50}$\checkmark$&\cellcolor{SeaGreen!50}$\checkmark$&&&\cellcolor{SeaGreen!50}$\checkmark$&\cellcolor{SeaGreen!50}$\checkmark$&\cellcolor{SeaGreen!50}$\checkmark$\\\hline
		\multicolumn{1}{ |c|}{S3}&\cellcolor{SeaGreen!50}$\checkmark$&\cellcolor{SeaGreen!50}$\checkmark$&\cellcolor{SeaGreen!50}$\checkmark$&&\cellcolor{SeaGreen!50}$\checkmark$&&\cellcolor{SeaGreen!50}$\checkmark$&\cellcolor{SeaGreen!50}$\checkmark$&\cellcolor{SeaGreen!50}$\checkmark$&\cellcolor{SeaGreen!50}$\checkmark$&&\cellcolor{SeaGreen!50}$\checkmark$&\cellcolor{SeaGreen!50}$\checkmark$\\\hline
		\multicolumn{1}{ |c|}{S4}&&&&&&&&&\cellcolor{SeaGreen!50}$\checkmark$&\cellcolor{SeaGreen!50}$\checkmark$&\cellcolor{SeaGreen!50}$\checkmark$&\cellcolor{SeaGreen!50}$\checkmark$&\\\hline
		\multicolumn{1}{ |c|}{S5}&\cellcolor{SeaGreen!50}$\checkmark$&\cellcolor{SeaGreen!50}$\checkmark$&\cellcolor{SeaGreen!50}$\checkmark$&\cellcolor{SeaGreen!50}$\checkmark$&\cellcolor{SeaGreen!50}$\checkmark$&\cellcolor{SeaGreen!50}$\checkmark$&\cellcolor{SeaGreen!50}$\checkmark$&\cellcolor{SeaGreen!50}$\checkmark$&\cellcolor{SeaGreen!50}$\checkmark$&\cellcolor{SeaGreen!50}$\checkmark$&\cellcolor{SeaGreen!50}$\checkmark$&\cellcolor{SeaGreen!50}$\checkmark$&\cellcolor{SeaGreen!50}$\checkmark$\\\hline
	\end{tabular}
	
\caption{Applicability of each improvement strategy to SSB-QB4OLAP queries.}
\label{tab.imp_app}

\end{table}

\begin{table}
\caption{TPC-H metrics: improvement evaluation}
\label{tab.tpcres}
\scriptsize
\begin{tabularx}{\linewidth}{@{\extracolsep{\fill}} |p{0.8cm}|p{0.8cm}|p{1.5cm}|p{1.5cm}|X|}
\hline
 & Power (QpH) & Throughput (QpH) & Composite (QpH) & Interval (sec) \\ \hline
Na\"{i}ve & 63.8 & 75.6 & 69.5 & 1237.6 \\ \hline
ES1 & 253.1 & 293.3 & 272.4 & 319.2 \\ \hline
ES2 & 402.4 & 361.2 & 381.2 & 259.1 \\ \hline
ES3  & 326.7 & 353.9 & 340.0 & 264.5 \\ \hline
ES6 & 354.5 & 108.3 & 196.0 & 864.2 \\ \hline
ES14 & 217.3 & 148.9 & 179.9 & 628.7 \\ \hline
ES15 & 257.4 & 198.7 & 226.2 & 471.0 \\ \hline
ES16 & 415.5 & 254.0 & 324.9 & 368.4 \\ \hline
ES7 & 706.8 & 561.9 & 630.2 & 166.6 \\ \hline
ES17 & 427.2 & 368.4 & 396.7 & 254.1 \\ \hline
ES18 & 427.6 & 339.4 & 381.0 & 275.8 \\ \hline
ES19  & 456.6 & 379.6 & 416.4 & 246.6 \\ \hline
ES4  & 375.8 & 215.9 & 284.9 & 433.4 \\ \hline
ES8 & 253.6 & 171.5 & 208.6 & 545.7 \\ \hline
ES9 & 227.0 & 146.5 & 182.4 & 638.8 \\ \hline
ES10 & 214.7 & 148.0 & 178.2 & 632.6 \\ \hline
ES5  & 490.8 & 418.6 & 453.3 & 223.6 \\ \hline
ES11 & 693.1 & 750.1 & 721.0 & 124.8 \\ \hline
ES12  & 472.4 & 368.9 & 417.5 & 253.7 \\ \hline
ES13  & 380.2 & 327.2 & 352.7 & 286.1 \\ \hline
\end{tabularx}
\end{table}

\subsubsection{Comparison with SSB-QB}
\label{sec:eval.setup.comp}

\begin{figure}[t]
  \begin{center}
    \fontsize{6}{6}\selectfont
     \includegraphics[width=\linewidth]{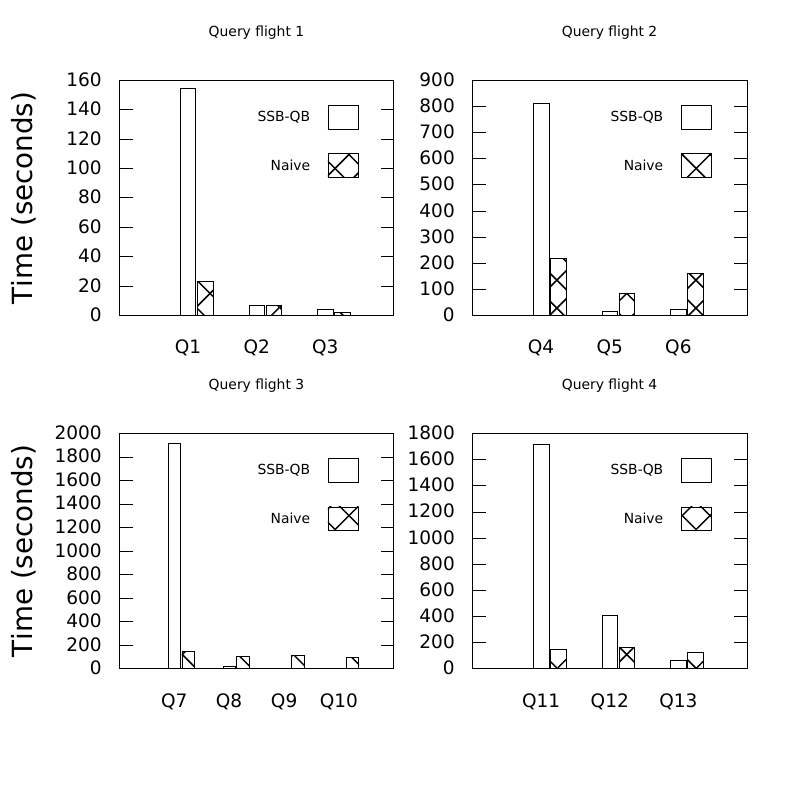}
     \vspace{-1cm}
    \caption{SSB-QB and Na\"{i}ve queries execution time}
    \label{graph:QB-Naive}
  \end{center}
\end{figure}

We also wanted to compare the queries produced by our 
na\"{i}ve approach, and the best and worst cases of the improved queries, against the SSB-QB queries. Thus, we implemented SSB-QB in our experimental setting, and ran the queries. 
Table \ref{tab.tpcalle} shows the results obtained for each TPC-H metric, and  
Figure \ref{graph:QB-Naive} presents a detailed comparison on the execution time for each query. We compare  SSB-QB best case (the minimum execution time) against the na\"{i}ve SSB-QB4OLAP worst case (the maximum execution time).

\begin{table}[t]
\caption{TPC-H metrics comparison}
\label{tab.tpcalle}
\scriptsize
\begin{tabularx}{\linewidth}{@{\extracolsep{\fill}} |p{1.7cm}|p{0.8cm}|p{1.1cm}|p{1.2cm}|X|}
\hline
 & Power (QpH) & Throughput (QpH) & Composite (QpH) & Interval (sec) \\ \hline
SSB-QB\cite{Kampgen2013} & 69.9 & 17.2 & 34.7 & 5447.0 \\ \hline
SSB-QB4OLAP Na\"{i}ve & 63.8 & 75.6 & 69.5 & 1237.6 \\ \hline
SSB-QB4OLAP ES14 (worst case)& 217.3 & 148.9 & 179.9 & 628.7 \\ \hline
SSB-QB4OLAP ES11 (best case)& 693.1 & 750.1 & 721.0 & 124.8 \\ \hline
\end{tabularx}

\end{table}

%\vspace{-1.cm}
\subsection{Discussion}
\label{sec:eval.discussion}
Regarding  the improvement scenarios, results show that, for 
  the TPC-H Composite metric, scenario ES11 outperforms the other ones,
 with a 10X improvement with respect to the na\"{i}ve scenario (see Table \ref{tab.tpcres}), and a 10X speed-up in the execution time for the query mix. The second best scenario is ES7, with a 9X improvement on TPC-H Composite with respect to the na\"{i}ve scenario and a 9X speed-up. However, the average execution time per query is similar in both scenarios,  except for queries Q7 (where ES7 outperforms  ES11) and Q12 (where ES11 outperforms   ES7). Both scenarios apply S1, S2, and S4 (with \ttt{VALUES} splitting of    \ttt{FILTER} conditions), but ES7 applies S3, while ES11 
applies S5 with OC1  reordering (Figure \ref{fig.evalscenarios}).

Regarding the impact of each improvement strategy (Table \ref{tab.tpcres}),   strategies S1 and S2 combined  yield a   5.5X improvement with respect to na\"{i}ve queries. 
However, we cannot be conclusive on the impact of strategy S3. Note that the pairs of scenarios (ES6,ES4) and (ES7,ES5) only differ on the application of this strategy. In the first case, the scenario where S3 is applied performs worse (ES6), while in the second case the scenario where S3 is applied performs better (ES7). 
For  S4, our results show that,  replacing \ttt{FILTER} disjunctive conditions with \ttt{VALUES} clauses, improves performance (ES3 vs. ES7 and ES2 vs. ES5), while   \ttt{UNION} downgrades the performance (ES3 vs ES6 and ES2 vs. ES4). Finally, we cannot be conclusive on the impact of reordering graph patterns. 

Comparing our approach with SSB-QB, although the values for TPC-H Power metric are very similar, values for TPC-H composite show that even our na\"{i}ve approach represents a 2X improvement with respect 
to SSB-QB (Table \ref{tab.tpcalle}).  Considering our less improved scenario (ES14), we get a 5X enhancement, and 20X if we consider our best improved scenario (ES11). 
A detailed analysis on the execution time of each query (see Figure \ref{graph:QB-Naive}) shows that our approach outperforms SSB-QB for   Q1, Q4, Q7, Q11, and Q12.

We next further discuss the reasons why our na\"ive approach performs  better than the SSB-QB queries.

\begin{itemize}
	\item SSB-QB queries include an ORDER BY clause to order results, while our queries do not. 
	\item  As a consequence of the absence of level attributes, SSB-QB queries use string comparison on IRIs to fix level members, while we can use comparison over other data types, like, for example,  numeric values. It is well-known that string comparison is usually slower that integer comparison.
	\item The BGPs used to traverse hierarchies in SSB-QB may not take advantage of Virtuoso indexes.
\end{itemize}

To illustrate the last point we first give some insight on Virtuoso, and then present an example. The 
Virtuoso triple store uses a relational database to store data. In particular, all the triples are stored in a single table with four columns named graph (G), subject (S), predicate (P), and object (O). Two full, and three partial indices are implemented\footnote{\url{http://virtuoso.openlinksw.com/dataspace/doc/dav/wiki/Main/VirtRDFPerformanceTuning}}: \\\\
%\begin{itemize}
- PSOG: primary key index \\
- POGS: bitmap index for lookups on object value.\\
- SP: partial index for cases where only S is specified.\\
- OP: partial index for cases where only O is specified.\\
- GS: partial index for cases where only G is specified.\\
%\end{itemize}

Since the primary key is PSOG, data are physically ordered on this criteria.
Our strategy takes advantage of this index, while SSB-QB does not.
As an example, consider Q8 from SSB-QB4OLAP.

\begin{querynonum}
\label{ex:q8}
\textbf{Q8:} Revenue volume for lineorder transactions by customer city, supplier city and year, for suppliers and clients within the United States, and transactions issued between 1992 and 1997.
\end{querynonum}

Figures \ref{q8-ssbqb} and \ref{q8-naive} present the SPARQL representation of Query \ref{ex:q8} according to SSB-QB and to our na\"ive approach, respectively.
In particular, notice the BGPs that implement the \textsc{Roll-up} operation over the \te{Time} dimension (lines 8-12 in Figure \ref{q8-ssbqb} and lines 8-13 in Figure \ref{q8-naive}): Even though our approach uses more BGPs, at the time of the evaluation of each BGP, only the object of the triple is unknown, while in SSB-QB, subjects are also unknown. 

\begin{figure}
\begin{lstlisting}[style=sparql]
SELECT ?c_city ?s_city ?d_year 
       sum(?rdfh_lo_revenue) as ?lo_revenue  
FROM <http://lod2.eu/schemas/rdfh-inst#ssb1_ttl_qb>  
FROM <http://lod2.eu/schemas/rdfh#ssb1_ttl_dsd> 
FROM <http://lod2.eu/schemas/rdfh#ssb1_ttl_levels> 
WHERE {  
	?obs qb:dataSet rdfh-inst:ds.  
	?obs rdfh:lo_orderdate  ?d_date. 
	?d_yearmonthnum skos:narrower  ?d_date.
	?d_yearmonth skos:narrower  ?d_yearmonthnum. 
	?d_year skos:narrower  ?d_yearmonth.  
	rdfh:lo_orderdateYearLevel skos:member  ?d_year. 
	?obs rdfh:lo_custkey ?c_customer.
	?c_city skos:narrower  ?c_customer.
	?c_nation skos:narrower  ?c_city.
	?c_region skos:narrower  ?c_nation.
	rdfh:lo_custkeyRegionLevel skos:member  ?c_region.
	?obs rdfh:lo_suppkey ?s_supplier.
	?s_city skos:narrower  ?s_supplier.
	?s_nation skos:narrower  ?s_city.
	?s_region skos:narrower  ?s_nation.
	rdfh:lo_suppkeyRegionLevel skos:member  ?s_region.
	?obs rdfh:lo_revenue ?rdfh_lo_revenue. 
	FILTER(?c_nation = rdfh:lo_custkeyNationUNITED-STATES ).
	FILTER(?s_nation = rdfh:lo_suppkeyNationUNITED-STATES ).
	FILTER(
	 str(?d_year) >= "http://lod2.eu/schemas/rdfh#
	 lo_orderdateYear1992" and 
	 str(?d_year) <= "http://lod2.eu/schemas/rdfh#
	 lo_orderdateYear1997").
} 
GROUP BY ?d_year ?c_city ?s_city 
ORDER BY ASC(?d_year) DESC(?lo_revenue)    
\end{lstlisting}
\caption{Query 8 (SSB-QB)}
\label{q8-ssbqb}
\end{figure}

\begin{figure}
\begin{lstlisting}[style=sparql]
 SELECT ?plm2 ?plm3 ?plm5  (SUM(xsd:float(?m4)) as ?ag1) 
FROM <http://www.fing.edu.uy/inco/cubes/instances/ssb_qb4olap> 
FROM <http://www.fing.edu.uy/inco/cubes/schemas/ssb_qb4olap> 
WHERE { 
	?o a qb:Observation .
	?o qb:dataSet rdfh-inst:ds .
	?o rdfh:lo_revenue ?m4 .
	?o rdfh:lo_orderdate ?lm1 .
	?lm1 qb4o:memberOf rdfh:lo_orderdate . 
	?lm1 schema:dateInMonth ?plm1 .
	?plm1 qb4o:memberOf schema:month . 
	?plm1 schema:monthInYear ?plm2 .  
	?plm2 qb4o:memberOf schema:year .
	?o rdfh:lo_custkey ?lm2 .
	?lm2 qb4o:memberOf rdfh:lo_custkey . 
	?lm2 schema:inCity ?plm3 .
	?plm3 qb4o:memberOf schema:city . 
	?plm3 schema:inNation ?plm4 .
	?plm4 qb4o:memberOf schema:nation .
	?o rdfh:lo_partkey ?lm3 .
	?o rdfh:lo_suppkey ?lm4 .
	?lm4 qb4o:memberOf rdfh:lo_suppkey . 
	?lm4 schema:inCity ?plm5 .
	?plm5 qb4o:memberOf schema:city . 
	?plm5 schema:inNation ?plm6 .
	?plm6 qb4o:memberOf schema:nation .
	?plm4 schema:nationName> ?plm41 .
	?plm6 schema:nationName> ?plm61 .
	?plm2 schema:yearNum ?plm21 .
	FILTER	(	?plm41 = "UNITED STATES"  &&
					?plm61 = "UNITED STATES"  &&
					?plm21 >= 1992  && ?plm21 <= 1997) 
}
GROUP BY ?plm2 ?plm3 ?plm5
\end{lstlisting}

\caption{Query 8 (SSB-QB4OLAP na\"ive)}
\label{q8-naive}
\end{figure}

\section{Related Work}
\label{sec:related}

We identify two major approaches in OLAP analysis of SW data. 
The first one consists in extracting MD data from the web, and loading them into traditional data management systems for OLAP analysis. 
This approach requires a local DW to store the extracted data, a restriction that clashes with the autonomous 
and highly volatile nature of web data sources. Relevant to this line of research, are the works by Nebot and 
Llavori~\cite{Nebot2012} and K\"{a}mpgen and Harth~\cite{Kampgen2011}. 
We will discuss here a different  line of work, which explores data models and tools that allow publishing and performing OLAP-like analysis directly over the SW, \textit{representing MD data in RDF}. 
This is closely related with the concepts of \textit{self-service BI}, which aims at incorporating web data into the 
decision-making process~\cite{DBLP:journals/jdwm/AbelloDEGMNPRTVV13}, and \textit{exploratory OLAP}~\cite{Abello2015}. 

Ibrahimov et al.~\cite{IbragimovHPZ14} present a framework for Exploratory BI over Linked Open Data.
Their goal is to semi-automatically derive MD schemas and instances, from already published Linked Data.
The proposed framework uses the QB4OLAP vocabulary to represent the discovered OLAP schemas, while the VoID vocabulary is used to link the schema with available SPARQL endpoints that can be used to populate it. Although the envisioned framework should be able to answer MDX queries, few details are provided on the translation process from MDX queries to SPARQL queries over QB4OLAP. Although expert OLAP users are likely to know MDX, in a self-service BI environment most users are not so proficient, in our opinion, 
we need a \textit{more intuitive language}, that can deal  only with cubes, an intuitive  data structure for most analytical users. 
%Further, the semantics of MDX is rather unclear, while 
%the one of \QL~is.  

Literature on MD data representation in RDF can be further organized in two categories: (i) Those that use specialized RDF vocabularies to explicitly define the data cubes;  and (ii) Those that implicitly define a data cube over existing RDF data graphs. Our work follows the explicit approach, and extends the QB vocabulary to include the MD structure. 
K\"{a}mpgen et al.~\cite{Kampgen2013,Kampgen2012} also attempt to override the lack of structure in QB 
defining an OLAP  data model  on top of QB and other  vocabularies. They use extensions to  the 
SKOS vocabulary\footnote{\url{http://www.w3.org/2011/gld/wiki/ISO_Extensions_to_SKOS}} 
to represent the hierarchical structure of the dimensions.  
In this representation,  levels can belong to only one hierarchy, and level attributes are not supported. 
In \cite{Kampgen2013} the authors implement some OLAP operators over those extended cubes, using SPARQL queries, 
restricted to data cubes with only one hierarchy per dimension. They also explore the use of RDF aggregate views to improve  performance. 
This approach  requires specialized OLAP engines for analytical queries over RDF data, 
instead of  traditional triple stores.

The WaRG project\footnote{\url{https://team.inria.fr/oak/projects/warg/}} proposes a new analytical model to implicitly define data cubes over RDF graphs. 
The core concepts are the Analytical Schema (AnS), a graph that represents an MD view over existing RDF data, 
following the classical Global-as-View data integration approach, and Analytical Queries (AnQ) over AnS, which can be implemented as SPARQL BGPs~\cite{Colazzo2014,AziraniGMR15}. Although they show how some OLAP operations can be implemented as AnQs, key operations like   
\textsc{Roll-up} are just briefly sketched.
Moreover, AnS does not support the definition of complex dimension hierarchies.

Regarding SPARQL query processing, many  works study the   complexity of query evaluation \cite{Perez2009,Schmidt2010}.
In \cite{letelier2013static} the authors focus on  
the static analysis of SPARQL queries, in particular those that contain the \texttt{OPTIONAL} operator.
%They propose to represent queries as pattern trees, and provide a set of transformation %rules over this      
%trees that can be used to obtain better queries.
Tsialimanis et. al \cite{tsialiamanis2012heuristics} propose a heuristic approach to the optimization for  
SPARQL joins, based on the selectivity of graph patterns. All of these   are general-purpose studies. 
On the contrary, \textit{we take advantage of the  characteristics of our data model} 
(e.g., the OLAP operators, 
and the information provided by QB4OLAP metadata) 
to   define optimization rules that  may not apply to a more  generic scenario. 
  
Jakobsen et al.~\cite{JakobsenAHP15} study the improvement of SPARQL queries over QB4OLAP data cubes. 
To reduce the number of joins (BGPs) needed to traverse hierarchies, 
they propose to generate denormalized representations of data instances called \textit{star patterns} and \textit{denormalized patterns}, which resemble
relational representation strategies for MD data. 
The idea behind this approach is to directly link facts (observations) with attribute values of related level members.
Although preliminary results show an improvement in queries performance, this approach prevents level members from being reused and referenced, breaking the Linked Data nature of QB4OLAP data instances. 

\section{Conclusion}
\label{sec:conclusion}

In this paper we proposed the use of a high-level  language (\QL) over data cubes, to express OLAP queries at a conceptual level. We showed that these queries can be automatically translated  into efficient SPARQL ones. For this, we first used the metadata provided by the QB4OLAP vocabulary to obtain a na\"{i}ve translation of \QL~ programs  to SPARQL queries, and then, we adapted general-purpose SPARQL optimization techniques to the OLAP setting, to obtain better performance. Our experiments over synthetic data (an adaptation of the Star-Schema TPC-H benchmark) showed that even the   na\"{i}ve approach outperforms other proposals, and suggest the best combinations of optimization strategies. An application to explore SW cubes, write, and execute \QL~queries, completes our contibutions. We believe that these results can encourage and promote the publication and sharing of MD data on the SW. 
We plan to continue working in this direction, extending
\QL~(and the corresponding translations) with other OLAP operations.

\appendix

\section*{Acknowledgments}  
Alejandro Vaisman was partially supported by 
PICT-2014 Project 0787, from the Argentinian Scientific Agency.

\section{Prefixes used in this paper}
Below, we show the prefixes  used in this paper.
\begin{figure}[!h]
\begin{lstlisting}[style=tiny]
PREFIX xsd: <http://www.w3.org/2001/XMLSchema#>
PREFIX qb:  <http://purl.org/linked-data/cube#>
PREFIX qb4o: <http://purl.org/qb4olap/cubes#>
PREFIX sdmxm: <http://purl.org/linked-data/sdmx/2009/measure#>
PREFIX sdmxd: <http://purl.org/linked-data/sdmx/2009/dimension#>
PREFIX pr: <http://eurostat.linked-statistics.org/property#>
PREFIX citizen: <http://eurostat.linked-statistics.org/dic/citizen#>
PREFIX geo: <http://eurostat.linked-statistics.org/dic/geo#>
PREFIX age: <http://eurostat.linked-statistics.org/dic/age#>
PREFIX sex: <http://eurostat.linked-statistics.org/dic/sex#>
PREFIX app: <http://eurostat.linked-statistics.org/dic/asyl_app#>
PREFIX dt: <http://eurostat.linked-statistics.org/data/>
PREFIX ds: <http://eurostat.linked-statistics.org/data/migr_asyappctzm#>
PREFIX loc-ins: <http://www.fing.edu.uy/cubes/instances/>
PREFIX loc-sch: <http://www.fing.edu.uy/cubes/schemas/>
PREFIX sc: <http://www.fing.edu.uy/cubes/schemas/migr_asyapp#>
PREFIX instances: <http://www.fing.edu.uy/cubes/instances/migr_asyapp>
PREFIX citDim: <http://www.fing.edu.uy/cubes/dims/migr_asyapp/citizen#>
PREFIX time: <http://purl.org/qb4olap/dimensions/time#201409>
\end{lstlisting}	
%\caption{RDF prefixes used in this work}
%\label{fig:prefixes}
\end{figure}

\section{QB4OLAP Representation of the Asylum Applications Data Cube}
\label{appendixb}
Below, we show  how the Eurostat data cube in our running example, 
  looks like in QB4OLAP. 
 Note that  the structure   is defined in terms of  dimension levels, which  
represent the granularity of the observations in the data set (i.e., these levels are the lowest levels in the dimension hierarchies).

\begin{lstlisting}[style=script]
sc:migr_asyapp rdf:type qb:DataStructureDefinition ;
  qb:component [ qb:measure sdmxm:obsValue ; 
                 qb4o:aggregateFunction qb4o:sum ] ;
  qb:component [ qb4o:level pr:age ;
                 qb4o:cardinality qb4o:ManyToOne ] ;
  qb:component [ qb4o:level sdmxd:refPeriod ;
                 qb4o:cardinality qb4o:ManyToOne ] ;
  qb:component [ qb4o:level pr:sex ; 
                 qb4o:cardinality qb4o:ManyToOne] ;
  qb:component [ qb4o:level pr:geo ; 
                 qb4o:cardinality qb4o:ManyToOne ] ;
  qb:component [ qb4o:level pr:citizen ; 
                 qb4o:cardinality qb4o:ManyToOne ] ;
  qb:component [ qb4o:level pr:asyl_app ; 
                 qb4o:cardinality qb4o:ManyToOne ] .

dt:migr_asyappctzm qb:structure sc:migr_asyappctzmQB4O;.
\end{lstlisting}

An observation (represented in QB4OLAP) corresponding to the schema above, is shown below. It corresponds 
to the first row of Table \ref{fig.cubeinstance}. 

\begin{lstlisting}[style=script]
ds:M,SY,F,Y18-34,NASY_APP,DE,2014M09 a qb:Observation ;
  pr:age age:Y18-34 ; 
  sdmxd:refPeriod time:2001409 ;  
  pr:sex sex:F ; 
  pr:geo geo:DE ; 
  pr:citizen citizen:SY ; 
  pr:asyl_app app:NASY_APP ; 
  sdmxm:obsValue 425 .
\end{lstlisting}
%\qed
 
Dimensions are represented in QB4OLAP as follows. 
We define  the citizenship dimension 
\ttt{sc:citDim} of Figure \ref{fig.concEuro}, and 
  the hierarchy  
\ttt{sc:citGeoHier}, also 
declaring its  levels \ttt{pr:citizen}
and \ttt{sc:continent}. Also,   
we associate  attributes with  levels, e.g.,   \ttt{sc:contName} with  \ttt{sc:continent}. 
Finally, the rollups and  hierarchy steps (i.e, parent-child relationships) are defined. 
 
\begin{lstlisting}[style=script]
# Dimension definition
sc:citDim a qb:DimensionProperty ;
  rdfs:label "Applicant citizenship dimension"@en ;
	  qb4o:hasHierarchy sc:citGeoHier, sc:citGovHier .
	  
# Hierarchy definition
sc:citGeoHier a qb4o:Hierarchy ; 
   rdfs:label "Applicant citizenship Geo Hierarchy"@en ;
   qb4o:inDimension sc:citDim ;
   qb4o:hasLevel pr:citizen, sc:continent .
   
# Base level
pr:citizen a qb4o:LevelProperty ; 
  rdfs:label "Country of citizenship"@en ;
  qb4o:hasAttribute sc:counName. 
sc:counName  a qb4o:LevelAttribute ;
  rdfs:label "Country name"@en ; rdfs:range xsd:string . 
  
#Upper hierarchy levels
sc:continent a qb4o:LevelProperty ; 
  rdfs:label "Continent"@en ;
  qb4o:hasAttribute sc:contName .
sc:contName  a qb4o:LevelAttribute ;
  rdfs:label "Continent name"@en ; rdfs:range xsd:string .
  
#rollup relationships   
sc:inContinent a qb4o:RollupProperty .
sc:hasGovType a qb4o:RollupProperty .
#hierarchy step
_:ih1 a qb4o:HierarchyStep ; 
  qb4o:inHierarchy sc:citGeoHier ;
  qb4o:childLevel pr:citizen ;
  qb4o:parentLevel sc:continent ;
  qb4o:pcCardinality qb4o:OneToMany ;
  qb4o:rollup sc:inContinen t.
\end{lstlisting} 
 %\qed
 
\textit{Level members} are represented as instances of the class 
\ttt{qb4o:LevelMember}, and attached to the levels they belong to via the property  
\ttt{qb4o:memberOf}, as shown next, using  the dimension members for dimension~\ttt{sc:citDim,} corresponding to  Syria. Note that, for attribute instances, we need to link
IRIs  representing  level members, with literals, corresponding 
to attribute values.
\begin{lstlisting}[style=script]
citizen:SY 
  qb4o:memberOf pr:citizen ;
  sc:counName "Syria"@en ;
  sc:inContinent citDim:AS ;
  sc:hasGovType dbpedia:Unitary_state .
  
citDim:AS
  qb4o:memberOf sc:continent ;
  sc:contName "Asia" .
  
dbpedia:Unitary_state
  qb4o:memberOf sc:governmentType ;
  sc:govName "Unitary state"@en .
\end{lstlisting}
% \qed

\section*{References}

\bibliography{jws2016}

\end{document}